\documentclass[aps,prb,twoside,twocolumn,showpacs,floatfix,10pt,showpacs,superscriptaddress]{revtex4-1}

\usepackage{graphicx}
\usepackage{amsmath}
\usepackage{times}

\newcommand{\fig}[1]{Fig.~\ref{#1}}
\newcommand{\tab}[1]{Tab.~\ref{#1}}
\newcommand{\sect}[1]{Sect.~\ref{#1}}
\newcommand{\eq}[1]{Eq.~(\ref{#1})}
\providecommand{\onlinecite}[1]{\cite{#1}}

\newcommand{\K}{\text{K}}

\newcommand{\atom}{\text{atom}}
\newcommand{\nm}{\text{nm}}
\newcommand{\R}{\text{R}}
\newcommand{\BCC}{\text{BCC}}
\newcommand{\Fe}{\text{Fe}}
\newcommand{\Cu}{\text{Cu}}
\renewcommand{\deg}{\ensuremath{^\circ}}
\renewcommand{\phi}{\ensuremath{\varphi}}
\renewcommand{\theta}{\ensuremath{\vartheta}}

\begin{document}

\title{
  Thermodynamic and mechanical properties of \\
  copper precipitates in $\alpha$-iron from atomistic simulations
}

\author{Paul Erhart}
\email{erhart@chalmers.se}
\affiliation{
  Chalmers University of Technology,
  Department of Applied Physics,
  Gothenburg, Sweden
}
\author{Jaime Marian}
\author{Babak Sadigh}
\affiliation{
  Lawrence Livermore National Laboratory,
  Condensed Matter and Materials Division,
  Livermore, California, USA
}

\begin{abstract}
Precipitate hardening is commonly used in materials science to control strength by acting on the number density, size distribution, and shape of solute precipitates in the hardened matrix. The Fe--Cu system has attracted much attention over the last several decades due to its technological importance as a model alloy for Cu steels. In spite of these efforts several aspects of its phase diagram remain unexplained. Here we use atomistic simulations to characterize the polymorphic phase diagram of Cu precipitates in body-centered cubic (BCC) Fe and establish a consistent link between their thermodynamic and mechanical properties in terms of thermal stability, shape, and strength.
The size at which Cu precipitates transform from BCC to a close-packed 9R structure is found to be strongly temperature dependent, ranging from approximately 4 nm in diameter ($\sim\!2,700\,\text{atoms}$) at 200\,K to about 8 nm ($\sim\!22,800\,\text{atoms}$) at 700\,K. These numbers are in very good agreement with the interpretation of experimental data given Monzen {\it et al.} [Phil. Mag. A {\bf 80}, 711 (2000)]. The strong temperature dependence originates from the entropic stabilization of BCC Cu, which is mechanically unstable as a bulk phase. While at high temperatures the transition exhibits first-order characteristics, the hysteresis, and thus the nucleation barrier, vanish at temperatures below approximately 300\,K. This behavior is explained in terms of the mutual cancellation of the energy differences between core and shell (wetting layer) regions of BCC and 9R nanoprecipitates, respectively. The proposed mechanism is not specific for the Fe--Cu system but could generally be observed in immiscible systems, whenever the minority component is unstable in the lattice structure of the host matrix.
Finally, we also study the interaction of precipitates with screw dislocations as a function of both structure and orientation. The results provide a coherent picture of precipitate strength that unifies previous calculations and experimental observations.
\end{abstract}

\pacs{
  62.20.fq, 
  64.70.kd, 
  64.70.Nd, 
  81.30.Hd  
}

\date{\today}

\maketitle

\section{Introduction}

Copper impurities contribute to high-temperature embrittlement in reactor pressure vessels (RPV) ferritic steels, which is a serious concern regarding lifetime extension of existing nuclear reactors. \cite{OdeLuc01} This behavior originates from the very small solubility of Cu in $\alpha$-Fe (body-centered cubic, BCC) and the correspondingly large driving force for precipitation. Copper has been observed to initially form small vacancy-rich clusters, which dissolve over time under typical RPV conditions, leaving embryonic Cu clusters that interact with internal dislocations. \cite{Ode83} The evolution of these nano-sized precipitates must be taken into account in embrittlement models for lifetime predictions of RPV steels. \cite{OdeLuc01}

Binary Fe--Cu alloys can be used to understand Cu precipitation and dislocation-precipitate interactions as prototypes for more complex RPV steels. \cite{NakHorOhn07} Based on a series of careful experimental studies on these alloys, three transformation stages have been identified \cite{OthJenSmi91, OthJenSmi94}: ({\it i}) Cu precipitates nucleate in the BCC structure, ({\it ii}) as they grow they undergo a martensitic phase transition to a multiply-twinned 9R structure, which ({\it iii}) eventually transforms ---presumably via diffusion--- into the 3R structure.
\footnote{
  The 3R structure is closely related to the face-centered cubic ground-state structure of Cu as it also exhibits a stacking sequence with threefold periodicity (see \sect{sect:results_BCC_9R}).
}
This sequence was determined using {\it post-mortem} transmission electron microscopy (TEM) investigations and selected area diffraction. \cite{OthJenSmi91, OthJenSmi94, HabJen96, MonJenSut00, LeeKimKim07}

Several molecular dynamics (MD) and statics simulation studies \cite{PhyForEng92, OseMikSer95, OseSer96, OseSer97, LudFarPed98, HuLiWat99, BlaAck01, Bou01} have been carried out to investigate the structural transitions of precipitates and to shed light on the underlying atomic mechanisms as such information is difficult to access via {\it post-mortem} experimental probes. A number of these studies have documented the shear instability associated with the aforementioned phase transformation that develops in BCC precipitates above a certain size. \cite{PhyForEng92, LudFarPed98, HuLiWat99, BlaAck01}
It has also been shown that vacancies bind strongly to copper precipitates, affect their stability \cite{OseSer97, BlaAck01} and play a role in the loss of coherency between precipitate and matrix \cite{OseSer97, BlaAck01, Bou01}.
Furthermore, motivated by the importance of Cu precipitates for the mechanical properties of $\alpha$-iron several investigations have addressed their interaction with dislocations, in particular in the case of coherent BCC precipitates. \cite{LudFarPed98, ShiChoKwo07, ShiKimJun08, BacOse09}

The objective of the present work is to obtain a comprehensive picture of the correlation between structure and size of Cu precipitates in $\alpha$-iron as a function of temperature, and to elucidate the interaction of dislocations with 9R precipitates. To this end, we first obtain the size-temperature phase diagram of Cu precipitates using a massively parallel hybrid molecular dynamics/Monte Carlo (MD/MC) algorithm. \cite{SadErhStu12} This investigation reveals a strong temperature dependence of the precipitate size at which the BCC--9R transition occurs, which originates from the entropic stabilization of BCC Cu. At high temperatures the character of the transformation is decidedly first-order as it exhibits a pronounced hysteresis and a large latent heat. In contrast, at low temperatures both hysteresis and nucleation barrier vanish. At a microscopic scale the transition, however, appears to remain discontinuous.

Regarding the dislocation--precipitate interaction, we find that the critical shear strength required for a dislocation to pass a precipitate changes discontinuously across the BCC--9R transition by approximately a factor of two and exhibits a marked dependence on orientation on the 9R side of the transition. These results have important implications for the interpretation of {\it post-mortem} experiments carried out near room temperature in connection with material structure and performance under RPV operating conditions at temperatures in excess of 560\,K. \cite{OdeLuc01}

The paper is organized as follows. Section~\ref{sect:methodology} describes our simulation methodology. Precipitate structure and transition mechanism are described in \sect{sect:results_bcc_9r_transition}. The thermodynamics of the transition are detailed in \sect{sect:results_thermodynamics}, followed by an analysis of the interaction of screw dislocations with BCC and 9R precipitates in \sect{sect:results_dislocations}. Finally, in \sect{sect:discussion} the results are discussed with respect to thermodynamic implications, their impact on the interpretation of experimental data, and plasticity.

\section{Methodology}
\label{sect:methodology}

The Fe--Cu binary alloy system was modeled using semi-empirical interatomic potentials of the embedded atom method type. For Cu and Fe potentials by Mishin {\it et al.} \cite{MisMehPap01} and Mendelev {\it et al.} (version v2 in Ref.~\onlinecite{MenHanSro03}) were used, respectively, while Fe--Cu interactions were described using the potential by Pasianot and Malerba. \cite{PasMal07} This combination of potentials provides an accurate description of the alloy phase diagram and yields very good agreement with density functional theory calculations for the energetics of small Cu clusters in BCC Fe. \cite{PasMal07}

Simulations were carried out using the hybrid MD/MC algorithm introduced in Refs.~\onlinecite{SadErhStu12} as implemented in the massively parallel MD code \textsc{lammps}. \cite{Pli95} The approach is based on the variance-constrained semi-grand-canonical (VC-SGC) ensemble, \cite{SadErhStu12, SadErh12} which generalizes the semi-grand-canonical (SGC) ensemble to allow stabilization of multiphase equilibria. Orthorhombic simulation cells were employed with cell vectors oriented along $[111]$, $[1\bar{1}0]$, and $[11\bar{2}]$. The majority of the simulations was based on simulation cells comprising $42\times 52\times 45$ repetitions of the unit cells corresponding to 786,240 atoms with additional simulations of smaller precipitate sizes at low temperatures using $31\times 38\times 33$ unit cells (310,992 atoms). Periodic boundary conditions were applied in all directions. The Nos\'e-Hoover thermostat and barostat \cite{FreSmi01} were employed to equilibrate the system at ambient pressure and temperatures between 300 and 700\,K. The equations of motion were integrated using a timestep of 2.5\,fs. Simulations were run for at least $1.6\times 10^6$ MD steps and ---if necessary for reaching equilibrium and/or collecting more statistics--- were extended to up to $4\times 10^6$ MD steps. Every 20 steps the MD simulation was interrupted to carry out a number of VC-SGC MC trial moves corresponding to 10\%\ of a full MC sweep. Each simulation run thus comprises between 8,000 and 20,000 attempts per particle to swap the atom type.

Parameters for the VC-SGC MC simulations were obtained as follows. First a series of SGC-MC simulations was carried out to determine the relation between relative chemical potential $\Delta\mu$ and concentration at temperatures between 300 and 700\,K. The average constraint parameter $\phi$ in the VC-SGC MC scheme \cite{SadErhStu12} was then taken as the $\Delta\mu$ value at which the solid solution of Cu in Fe became unstable. The variance constraint parameter $\kappa$ was set to $2000\,\text{eV}/\atom/k_B T$.
\footnote{
  It was shown in Refs.~\onlinecite{SadErhStu12, SadErh12} that the range of reasonable values for $\kappa$ extends over several orders of magnitude, rendering this choice uncritical.
}
Finally, the target concentration $c_0$ was varied to obtain the desired total number of Cu atoms in the simulation cell and thereby the size of the Cu precipitate.

Simulations of dislocation-precipitate interactions were conducted at a temperature of 300\,K with open boundary conditions perpendicular to the shear plane. Shear stress was applied so as to produce a Peach-K\"ohler force on the $(\overline{1}10)_\BCC$ plane that intersects the Cu precipitate through its equatorial plane.

Structural changes in the simulations were detected using the Ackland-Jones parameter, \cite{AckJon06} which uses bond angle distributions to classify local environments as body-centered cubic (BCC), face-centered cubic (FCC), hexagonal-close-packed (HCP), or icosahedral. The reliability of the Ackland-Jones analysis was enhanced by position-averaging over blocks of 800 MD steps. Precipitate structures were analyzed and visualized using \textsc{OVITO}. \cite{Stu10}

\section{Structural aspects of BCC--9R transition}
\label{sect:results_bcc_9r_transition}

\subsection{General features of BCC and 9R precipitates}
\label{sect:results_BCC_9R}

\begin{figure*}
  \centering
\includegraphics[width=0.95\linewidth]{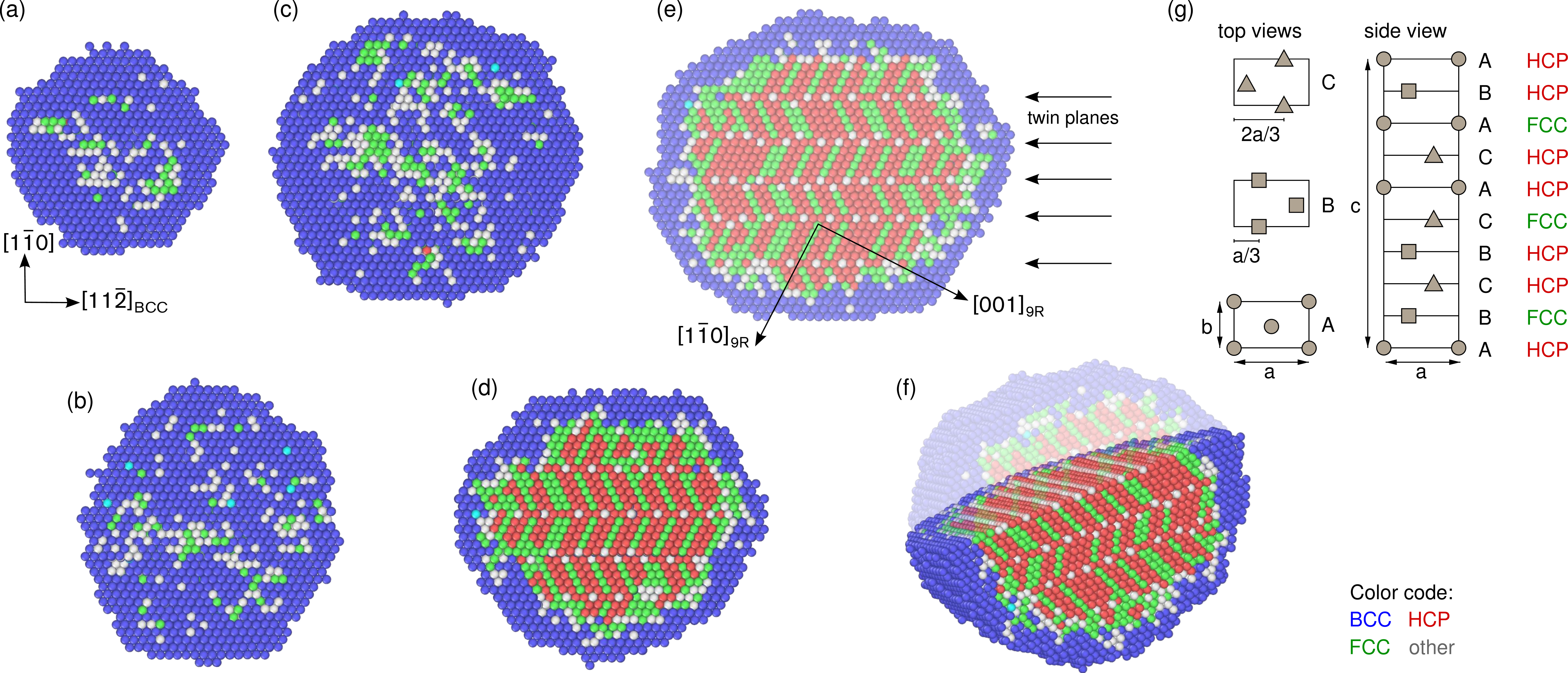}
  \caption{
    Cross sectional views of Cu precipitates obtained in VC-SGC MC/MD simulations at 700\,K exhibiting both (a-c) BCC and (d-f) 9R structures.
    The precipitates contain (a) 7,000, (b) 14,900, (c,d) 22,800, and (e,f) 34,600 atoms.
    The coloring indicates different local environments: blue=BCC, red=FCC, green=HCP, white=other.
    The 9R structure appears as a sequence of two HCP atoms and one FCC atom along the $[001]_{\text{9R}}$ direction [also compare (g)]. The twin planes are parallel to $\left\{110\right\}_\BCC$ planes and as can be seen in (f) extend through the entire precipitate. Note that the 9R cores of the precipitates shown in (d-f) are enclosed by a BCC Cu wetting layer. The out-of-plane vectors in (a-e) are $[111]_{\BCC}$ and $[\bar{1}10]_{\text{9R}}$, respectively.
    (g) Schematic of 9R structure adapted from Ref.~\onlinecite{OthJenSmi91}. Triangles, squares, and circles mark atomic positions in alternating close-packed planes, top views of which are shown on the left-hand side. The stacking sequence is shown on the right-hand side, which demonstrates that if only next nearest neighbors are considered (as in the Ackland-Jones analysis) the 9R unit cell is equivalent to three repetitions of the sequence HCP|FCC|HCP. In general, the 9R structure has monoclinic symmetry. For simplicity, here we show only the orthorhombic case.
  }
  \label{fig:viz_prec}
\end{figure*}

Two different types of precipitates were observed in the simulations, examples for which are shown in \fig{fig:viz_prec}. The precipitate shown in \fig{fig:viz_prec}(b) represents a snapshot from a VC-SGC MC/MD simulation at 700\,K, in which the total concentration was fixed at 2\%, a value significantly above the bulk solubility.
\footnote{
  The potentials employed here predict a solubility at 700\,K of approximately 0.06\%. \cite{PasMal07} This value corresponds to the stability of the BCC Fe solid solution with respect to BCC Cu. Note that in order to obtain the solubility shown in standard phase diagrams one needs to consider the free energy balance between solid solutions of BCC Fe and FCC Cu. \cite{CarCarLop06a} The difference is, however, small and therefore neglected here.
}
This oversaturation leads to formation of a spherical Cu precipitate containing approximately 14,900 atoms, which is practically Fe free. On average the precipitate is found to retain the BCC structure of the surrounding matrix.

The precipitate shown in \fig{fig:viz_prec}(e) was also obtained at 700\,K using, however, a target concentration of 4.5\%. It exhibits an elongated ellipsoidal shape, contains about 34,600 atoms and appears to be composed of ordered stacks of FCC and HCP atoms, which can be identified as belonging to a multiply-twinned 9R structure as illustrated in \fig{fig:viz_prec}(g).
\footnote{
  The 9R structure has also been observed e.g., at grain boundaries in silver \cite{ErnFinHof92} and gold \cite{MedFoiCoh01} as well as in single crystal copper wires. \cite{CheYanFan09}
}
There are three distinct ways in which close-packed planes can be stacked denoted here by A, B, and C. For example, the FCC structure is described by the stacking sequence ABC|ABC whereas the sequence AB|AB yields the HCP structure. By contrast the unit cell of the 9R structure comprises nine close-packed planes. Since the Ackland-Jones parameter considers only nearest neighbors it is suitable for distinguishing HCP and FCC environments but cannot directly identify the 9R structure. As shown in \fig{fig:viz_prec}(g) the 9R structure rather appears as a sequence of FCC and HCP atoms. Comparison of the ideal 9R structure in \fig{fig:viz_prec}(g) with the precipitates in \fig{fig:viz_prec}(d-f) demonstrates that the latter have adopted a multiply-twinned 9R structure.

It is important to note that 9R precipitates regardless of size and temperature feature a BCC Cu wetting layer between the BCC Fe matrix and the 9R Cu core. As will be described below, the latter plays a crucial role in understanding the thermodynamics of the BCC--9R transition as well as the dislocation-precipitate interaction.

The simulation results are substantiated by the observation that the structures of 9R precipitates shown in \fig{fig:viz_prec} closely resemble TEM micrographs of 9R Cu-precipitates in $\alpha$-Fe. \cite{OthJenSmi91, OthJenSmi94, LeeKimKim07} A more detailed analysis of the structure and shape of both BCC and 9R precipitates is deferred to Sects.~\ref{sect:results_BCC} and \ref{sect:results_9R}.

\begin{figure}
  \centering
\includegraphics[width=0.95\linewidth]{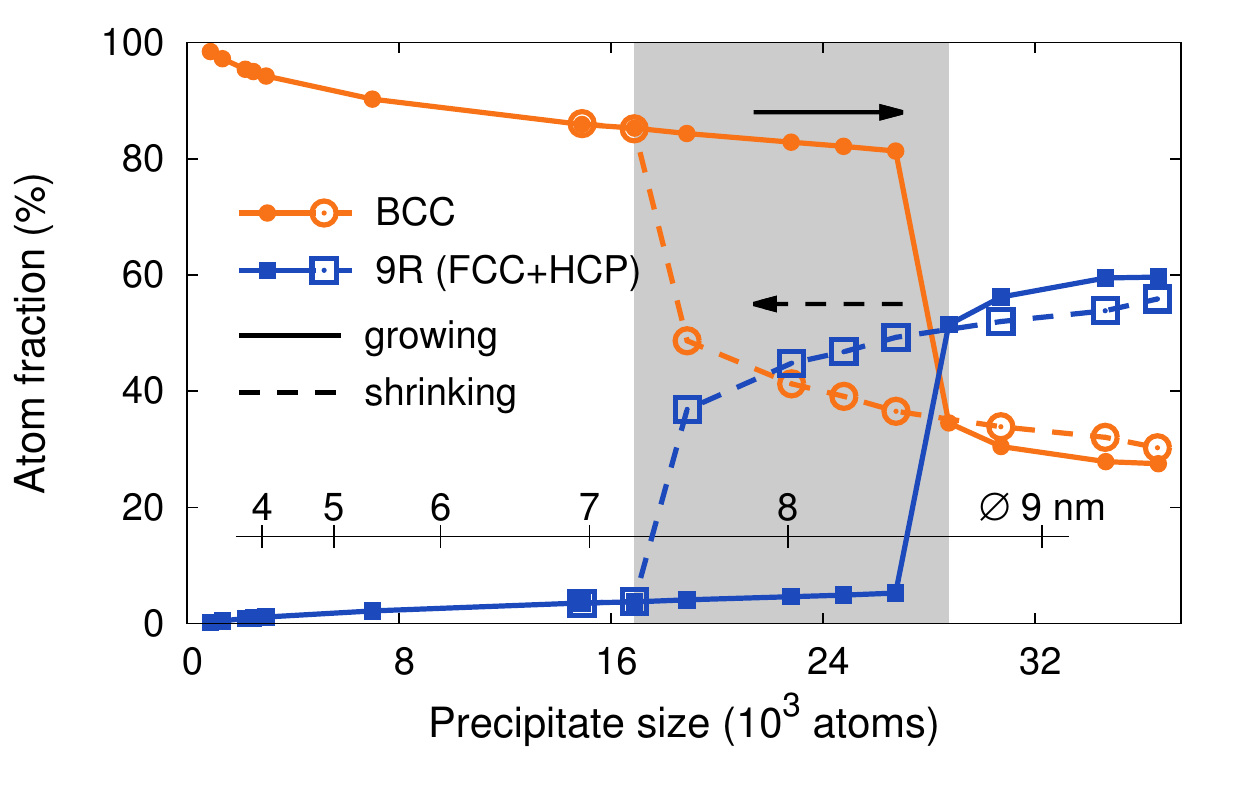}
  \caption{
    Fraction of atoms in Cu precipitate in different atomic neighborhoods at 700\,K At smaller sizes BCC motifs dominate indicating that Cu precipitates assume the BCC structure. For larger precipitates the number of atoms in close-packed environments (9R=FCC+HCP) is larger as Cu precipitates are now adopting a multiply-twinned 9R structure.
  }
  \label{fig:trans_struct}
\end{figure}

Next we consider the size and temperature dependence of the BCC--9R transition. Figure~\ref{fig:trans_struct} shows the number of atoms in different local environments as a function of precipitate size at 700\,K. Precipitates with more than about 28,000 atoms undergo a sharp structural transition as they transform from BCC to multiply-twinned 9R structures. 

The simulations also allow us to study the reverse transition. To explore this aspect, a series of simulations with decreasing Cu content was carried out starting from a 9R precipitate obtained in one of the earlier simulations. The results of this investigation are shown by the open symbols in \fig{fig:trans_struct}, which reveals that the reverse transition occurs for a precipitate size of approximately 18,000 atoms. There is thus a pronounced hysteresis between forward (BCC$\rightarrow$9R) and backward (9R$\rightarrow$BCC) transition extending from about 18,000 to 28,000 atoms. This is further illustrated by the BCC and 9R precipitates in Figs.~\ref{fig:viz_prec}(c) and (d), respectively, both of which contain about 22,800 atoms.
It should be stressed that the transition can be observed on the time scale of our simulations because it is martensitic (non-diffusive) in nature and transformations are observed when one structure becomes mechanically unstable with respect to the other.

\begin{figure}
  \centering
\includegraphics[width=0.95\linewidth]{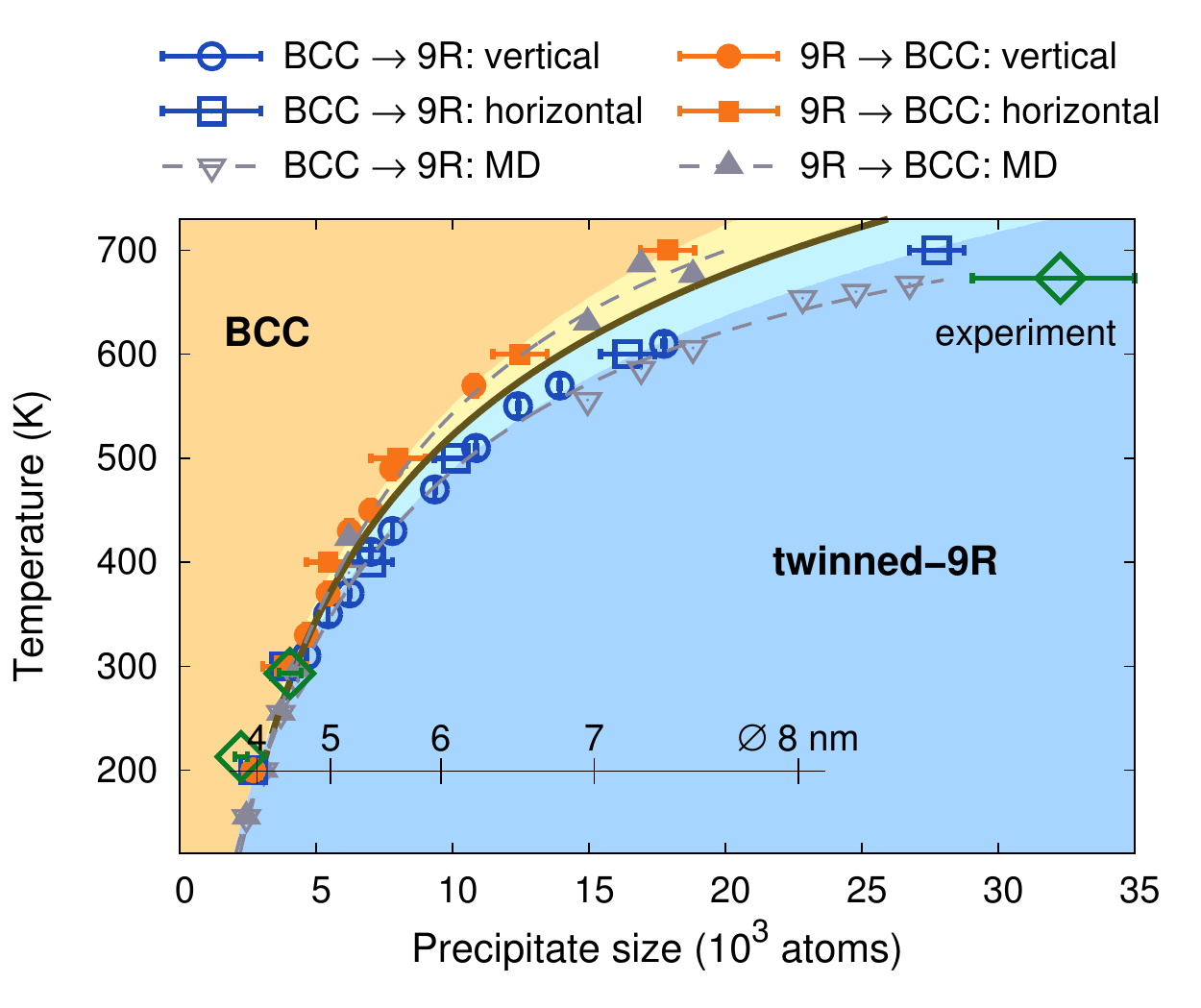}
  \caption{
    Phase diagram indicating the structure of Cu precipitates as a function of temperature and size. The light colored regions next to the dark gray phase transition line indicate the extent of the hysteresis, which vanishes below 300\,K. ``Vertical'' and ``horizontal'' data points have been obtained from VCSGC-MC/MD simulations at constant temperature and concentration, respectively (see \fig{fig:trans_struct}). The ``MD'' data points were obtained from conventional MD simulations (without atom type swaps) during cooling and heating, respectively (see \sect{sect:md_quenching}). The open diamonds indicate experimental data points from Ref.~\onlinecite{MonJenSut00}.
  }
  \label{fig:temp_trans}
\end{figure}

This type of analysis was carried out in 100\,K intervals between 200 and 700\,K, the results of which are shown by the ``horizontal'' data points in \fig{fig:temp_trans}. Further VCSGC-MC/MD simulations were then conducted at constant concentration for both decreasing and increasing temperature starting from previously equilibrated BCC and 9R precipitates, respectively. The results obtained in this fashion are marked as ``vertical'' data points in \fig{fig:temp_trans}, and are found to be consistent with the results from sampling at constant temperature. The two most important features of \fig{fig:temp_trans} are: ({\it i}) the precipitate size at which the BCC--9R transitions occurs decreases significantly with temperature and ({\it ii}) the width of the hysteresis between forward and backward transitions systematically decreases upon cooling and vanishes between 200 and 300\,K.

The pronounced hysteresis at high temperatures evident in \fig{fig:temp_trans} clearly indicates a first-order phase transition. The vanishing hysteresis with decreasing temperature, however, suggests that the first-order character diminishes and eventually the order of the transition changes. The thermodynamics of the transition as well as the strong temperature dependence of the critical transition size are investigated in detail in \sect{sect:results_thermodynamics}.

\subsection{Structure and shape of BCC precipitates}
\label{sect:results_BCC}

\begin{figure}
  \centering
\includegraphics[width=0.95\linewidth]{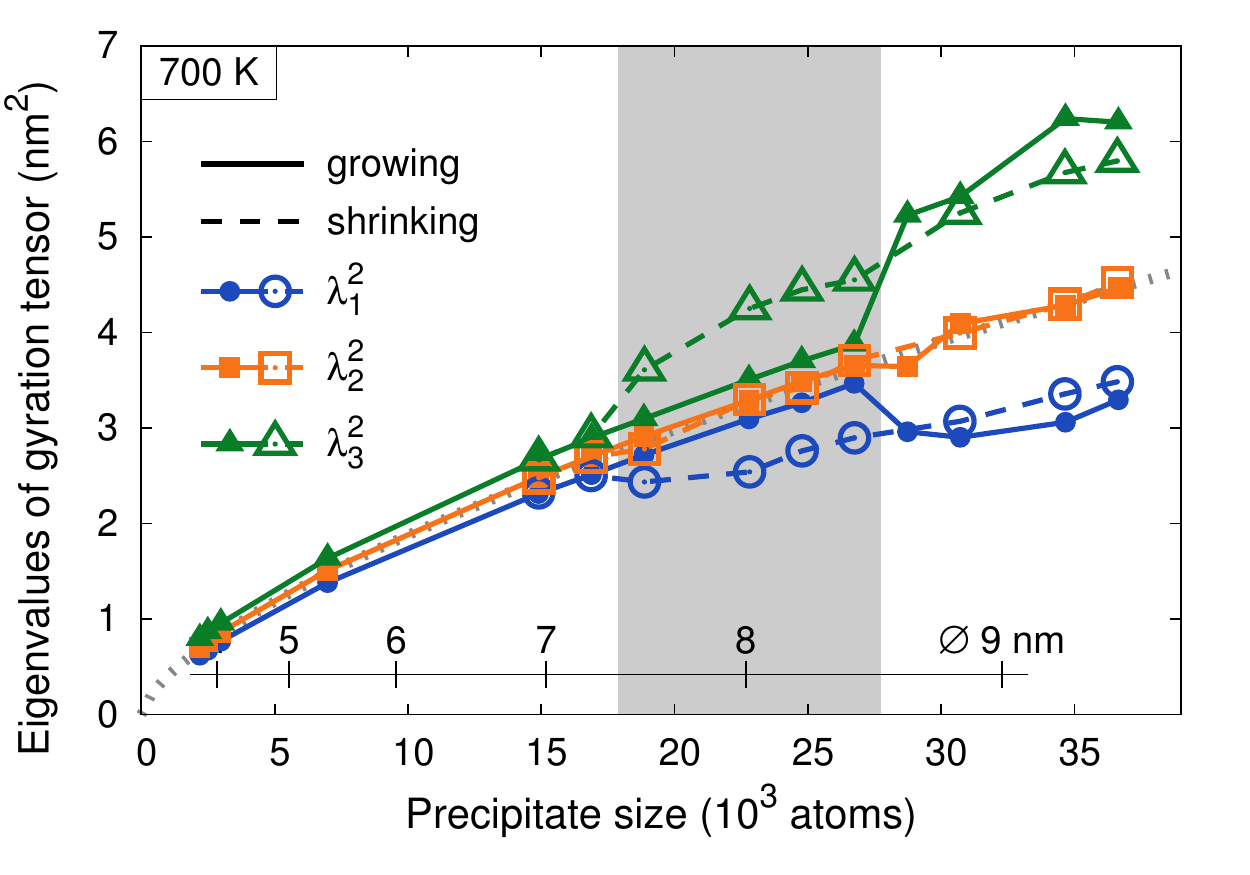}
  \caption{
    Eigenvalues of gyration tensor for Cu precipitates at 700\,K. The eigenvalues are denoted by $\lambda_i^2$ with $\lambda_1^2>\lambda_2^2>\lambda_3^2$.
  }
  \label{fig:trans_shape}
\end{figure}

As indicated in the previous section, BCC precipitates assume a spherical shape, while 9R precipitates tend to be ellipsoidal. To make this statement more quantitative, the eigenvalues of the gyration tensor have been computed, which is defined as
\begin{align}
  R_{\alpha\beta} &= \frac{1}{N} \sum_i^N r_{\alpha}^{(i)} r_{\beta}^{(i)},
\end{align}
where $N$ is the number of atoms in the precipitate, $r_{\alpha}^{(i)}$ denotes the component of the position vector of atom $i$ along Cartesian direction $\alpha$ and the origin of the coordinate system has been chosen such that $\sum_i^N r_{\alpha}^{(i)}=0$. Approximately speaking, the gyration tensor describes the ellipsoid that most closely resembles the point cloud $\{r_{\alpha}^{(i)}\}$. If all eigenvalues are identical ($\lambda_1^2=\lambda_2^2=\lambda_3^2$) the point cloud (or in the present case the precipitate) is best described by a sphere. On the other hand, for a prolate ellipsoid one would obtain one eigenvalue to be significantly larger than the other two ($\lambda_1^2>\lambda_2^2=\lambda_3^2$).

\begin{figure}
  \centering
\includegraphics[width=\linewidth]{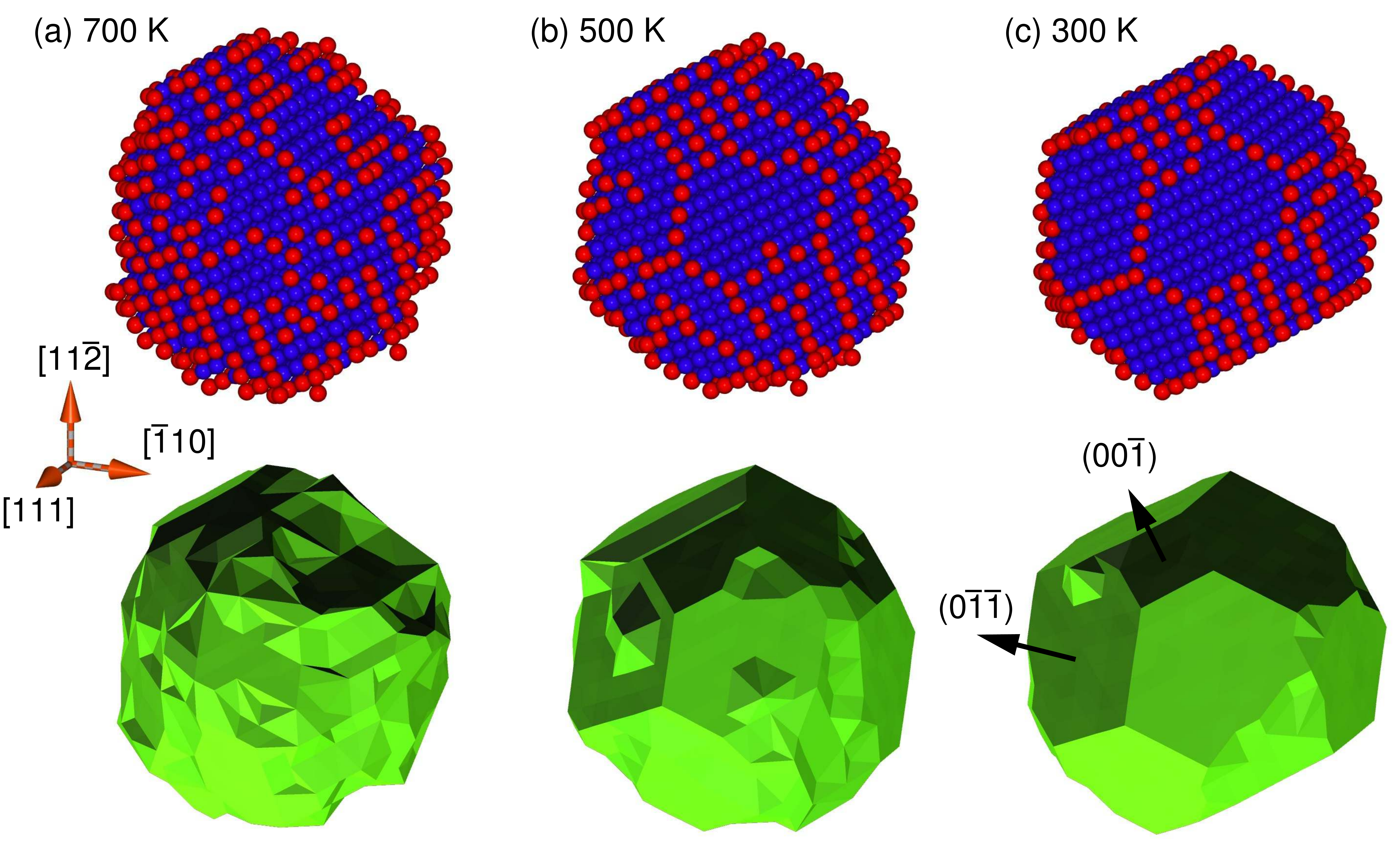}
  \caption{
    BCC Cu precipitates containing about 2,200 atoms that are representative for temperatures of (a) 700\,K, (b) 500\,K, and (c) 300\,K illustrating the increasing faceting with decreasing temperature. The top row shows the atomic positions where blue atoms have BCC coordination while red atoms indicate under-coordinated atoms at surfaces ledges and corners. The bottom row shows the convex hulls.
  }
  \label{fig:BCC_overview}
\end{figure}

Results of the gyration tensor analysis are shown in \fig{fig:trans_shape} for a temperature of 700\,K. For BCC precipitates the three eigenvalues are identical within the error bars, which is consistent with the spherical shape observable in \fig{fig:viz_prec}(a-c).

A visual inspection of the precipitates reveals a strong tendency for faceting as temperature is lowered. While at 700\,K precipitates are almost perfectly spherical, at 300\,K they exhibit strong faceting toward $\{110\}$ surfaces with small patches of $\{100\}$ facets

In \fig{fig:viz_prec}(a) time-averaged positions were employed to carry out the Ackland-Jones analysis, which reveals that the entire precipitate has adopted the BCC structure. If instead of using time-averaged positions, one employs instantaneous positions, FCC and HCP motifs can, however, be found as well. This observation is not unexpected since it is well known that the phonon dispersion of BCC Cu exhibits unstable modes. \cite{KraMarMet93, AckBacCal97, LiuWalGho05} MD simulations confirm that bulk BCC Cu remains unstable also at elevated temperature and over a wide range of volumes. This suggests that the stabilization of BCC Cu for precipitates with several thousand atoms is intimately connected to their finite size as well as the Fe--Cu interface. Indeed, one observes that for thin layers of BCC Cu wedged in between BCC Fe the interface tension is sufficient to stabilize BCC Cu at zero K. However, as the thickness of the Cu slab increases, the BCC structure again becomes unstable. This implies that the stabilization of BCC Cu precipitates with up to 28,000 atoms at 700\,K is the result of the combined effect of vibrations, finite size {\it and} interfaces that individually would not be sufficient to stabilize BCC Cu beyond very small sizes.

\subsection{Structure and shape of 9R precipitates}
\label{sect:results_9R}

\begin{figure*}
  \centering
\includegraphics[width=0.95\linewidth]{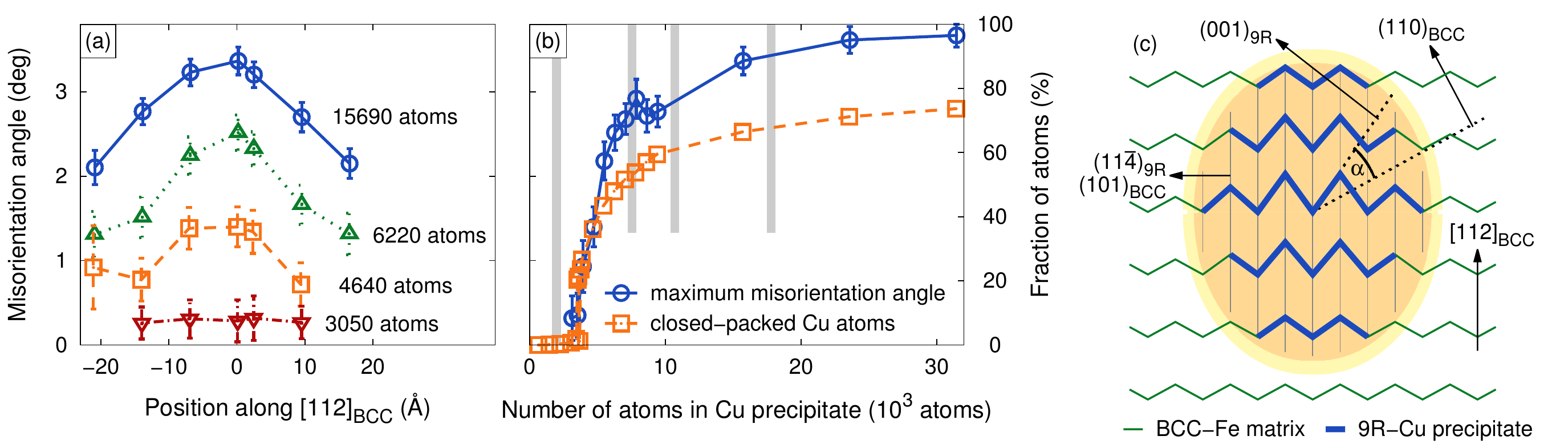}
  \caption{
    (a) Misorientation angle within a single 9R domain parallel to the twin plan for precipitates of various sizes at 300\,K.
    (b) Corresponding variation of the maximum misorientation angle with precipitate size. Also shown is the fraction of close-packed atoms in the precipitate. The vertical gray bars indicate the precipitate sizes at which dislocation simulations were carried out.
    (c) Schematic depiction of spatial variation of misorientation angle throughout a multiply-twinned 9R precipitate. Misorientation angles are strongly exaggerated for illustration.
  }
  \label{fig:angle_trans}
  \label{fig:schematic_9R_displ}
\end{figure*}

Next, we analyze the structure of 9R precipitates in the context of experimental observations and compare the 9R precipitates shown in \fig{fig:viz_prec}(d-f) with TEM images from e.g., Refs.~\onlinecite{OthJenSmi91, OthJenSmi94, LeeKimKim07}. In all cases precipitates exhibit the herringbone pattern that is characteristic of multiply-twinned 9R structures. \cite{OthJenSmi91, OthJenSmi94} Both simulated and experimentally observed precipitates are notably elongated in the direction parallel to the trace of the twin plane with aspect ratios of 1.2 (simulation) and 1.1--1.3 (experiment).
\footnote{
  For consistency with experiments, precipitate dimensions were measured parallel and perpendicular to the twin plane trace including only the 9R domains (that is excluding the Cu BCC shell, see \fig{fig:viz_prec}). In this fashion, one obtains aspect ratios of 1.1, 1.3, and 1.3 for the experimental precipitates in Fig.~2 of Ref.~\onlinecite{OthJenSmi91}, Fig.~4 of Ref.~\onlinecite{OthJenSmi94}, and Fig.~3 of Ref.~\onlinecite{LeeKimKim07}, respectively.
}
The orientation relations between BCC matrix and 9R domains observed in our simulations agree with experiment as well as theoretical predictions by Kajiwara \cite{Kaj76} with the  $[\bar{1}10]_{9\R}$ direction being parallel to $[111]_{\BCC}$ and the $(11\bar{4})_{9\R}$ plane being parallel to $(011)_{\BCC}$. Twinning occurs with respect to $\{11\bar{4}\}_{9\R}$ planes which are parallel to $\{110\}_{\BCC}$ planes.

Based on extensive TEM work, Othen {\it et al.} conducted a careful analysis of 9R variant thickness (twin spacing). Up to a diameter of about 13\,nm ({\it i.e.}, well above the size range covered by our simulations) they found twin domains to be evenly spaced with a width of about $1.5\pm 0.5\,\nm$. With the exception of very small precipitates that contained only two segments, they also found a slight increase of twin spacing with precipitate size. 

In general, the present simulations agree well with these observations. Specifically, we obtain twin spacings between $0.9\pm 0.1\,\nm$ at 300\,K and $1.2\pm 0.1\,\nm$ at 700\,K. Twins are evenly spaced and usually contain several structural defects such as stacking faults and steps inside the twin planes, very similar to the precipitates observed experimentally. \cite{OthJenSmi94}

An important structural parameter is the misorientation angle between $\{001\}_{9\R}$ and $\{110\}_{\BCC}$ planes [see \fig{fig:schematic_9R_displ}(c) for an illustration]. Using the angular deviation of $(003)_{9\R}$ and $(006)_{9\R}$ spots from $(110)_{\BCC,\Fe}$ spots in their dark field images, Othen and co-workers obtained an experimental value of $4\pm0.5\deg$ while from fringes in their bright field images they extracted a value of $4.3\pm0.7\deg$. Misorientation angles were extracted from the present simulations using time averaged atomic coordinates. In the largest precipitates misorientation angles in the center of the precipitates are just under $4\deg$, in reasonable agreement with experiment. A more detailed look at the results, however, reveals a more complex behavior than is suggested by the experimental analysis. Figure~\ref{fig:angle_trans}(a) demonstrates that misorientation angles vary across the Cu precipitate. They reach maximal values near the precipitate center and drop toward the precipitate-matrix interface. These features are shown in exaggerated fashion in \fig{fig:schematic_9R_displ}(c).

The maximum misorientation angle is practically temperature independent. It does, however, depend on precipitate size. Figure~\ref{fig:angle_trans}(b) shows that the maximum misorientation angle reaches a limiting value for large precipitates (about 3.6\deg\ for $N\gtrsim 10^4$ atoms) and almost completely smoothly approaches zero as the precipitate decreases in size. As in the case of the internal energy [compare \fig{fig:epot_trans}(a) below] there is a small discontinuity left at this temperature.

We conclude the structural description of 9R precipitates with an analysis of their shape in terms of the principal values of the gyration tensor as shown in \fig{fig:trans_shape}. Unlike BCC precipitates, 9R precipitates display a very pronounced asymmetry closely related to the orientation relations that govern BCC--9R interfaces. Most noticeably, 9R precipitates exhibit a pronounced elongation along $[1\bar{1}0]_{9\R}$, which is perpendicular to the figure plane in \fig{fig:viz_prec}(d,e). The other two principal axes are $[221]_{9\R}$ and $[11\bar{4}]_{9\R}$ with the precipitates being slightly longer along the former. As precipitate size increases the aspect ratio between the primary axis and the two shorter axes increases. Finally, similar to BCC precipitates, one observes a pronounced faceting of 9R precipitates as temperature is reduced with the dominant interfaces being parallel to $\{110\}_{\BCC}$ planes.

\subsection{Transition mechanism from BCC to 9R}
\label{sect:results_transmech}

\begin{figure}
  \centering
\includegraphics[width=0.95\linewidth]{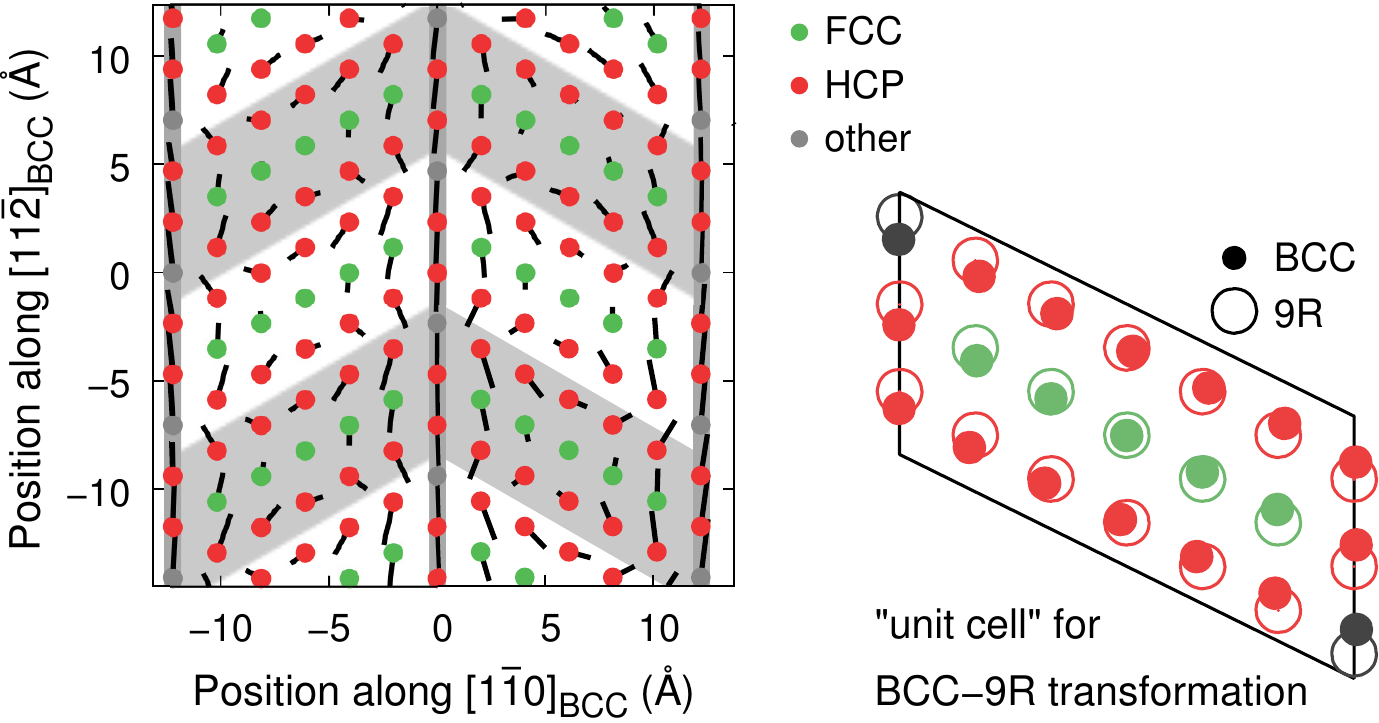}
  \caption{
    Atomic displacement pattern connecting BCC to multiply-twinned 9R structure. On the left hand side, a close up view of a slab (three atomic layers thick) is shown that is parallel to $(111)_{\BCC}$ and $(1\bar{1}0)_{9\R}$. Filled circles represent ideal BCC positions with the arrows indicating displacements that lead to 9R (scaled by a factor of three for clarity). Vertical gray bars mark twin planes while the slanted gray stripes are parallel to $\{001\}_{9\R}$ planes. On the right hand side, the smallest unit cell is shown that comprises all distinct displacements that take place during the BCC--9R transition leading to a single 9R domain. Color coding as in \fig{fig:viz_prec}.
  }
  \label{fig:transpath}
\end{figure}

The results presented in the previous sections already provide ample evidence for the martensitic character of the BCC--9R transition. To complete this analysis, we now describe the atomic displacement pattern associated with the BCC--9R transition. Figure~\ref{fig:transpath} shows a transformed region in a multiply-twinned 9R precipitate projected onto $(111)_\BCC$. The atomic positions correspond to ideal BCC sites with arrows indicating displacements that lead to 9R. The plot reveals a curl-like displacement pattern with a periodicity of three atomic layers along $[001]_{9\R}$. The right hand side of \fig{fig:transpath} shows the smallest unit cell that contains all distinct displacements in a single 9R domain. The cell contains 18 atoms and is three atomic layers thick along the $[1\bar{1}0]_{9\R}$ direction (perpendicular to the paper plane). The atoms also undergo regular displacements along $[1\bar{1}0]_{9\R}$, which are not visualized here. This path is specific for a twin spacing of about 1.2\,nm. Other twin spacings arise from qualitatively similar curl-like displacement patterns.

A visual analysis of the simulation results provides the following picture for the formation of multiply-twinned 9R particles. At temperatures at which the transition exhibits clear first-order character ($\gtrsim\,400\,\text{K}$), just prior to the transition {\em single} 9R domains appear that can extend over significant parts of the precipitate along $[11\bar{2}]_{\BCC}$ and $[111]_{\BCC}$ (compare \fig{fig:viz_prec}) but include only a couple of atomic layers along $[1\bar{1}0]_{\BCC}$. These domains are, however, unstable and decay rapidly. In contrast, right at the transition, the nucleation of the first 9R domain is immediately followed by the nucleation of a second 9R domain of opposite orientation. This event is crucial for stabilizing the first singly-twinned 9R nucleus. Subsequently, 9R segments are added rapidly as the transformation sweeps through the entire precipitate.

\begin{figure}
  \centering
\includegraphics[width=0.52\columnwidth]{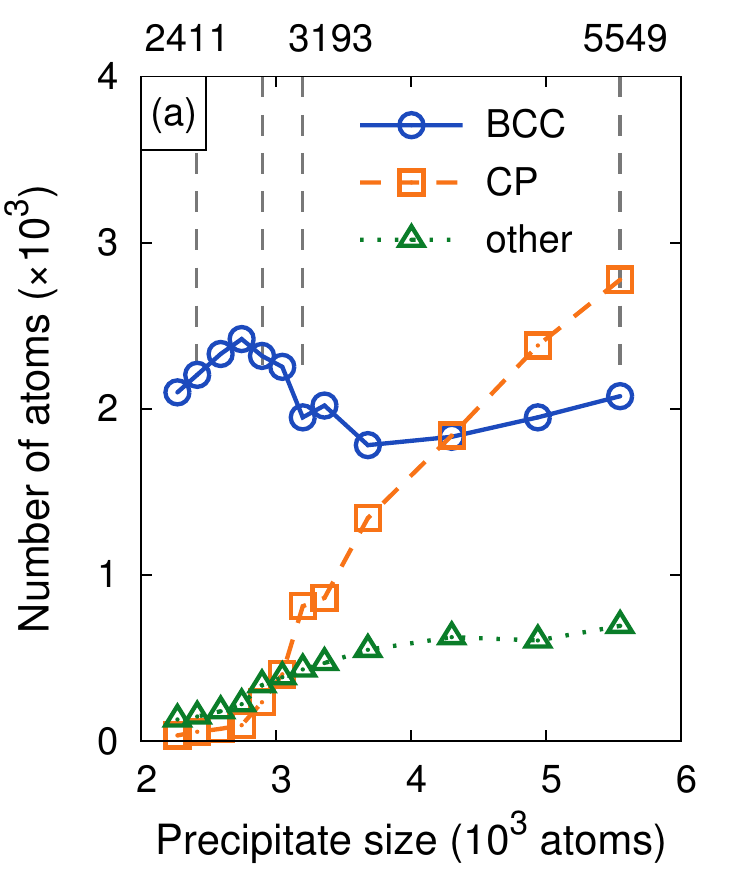}
\includegraphics[width=0.47\columnwidth]{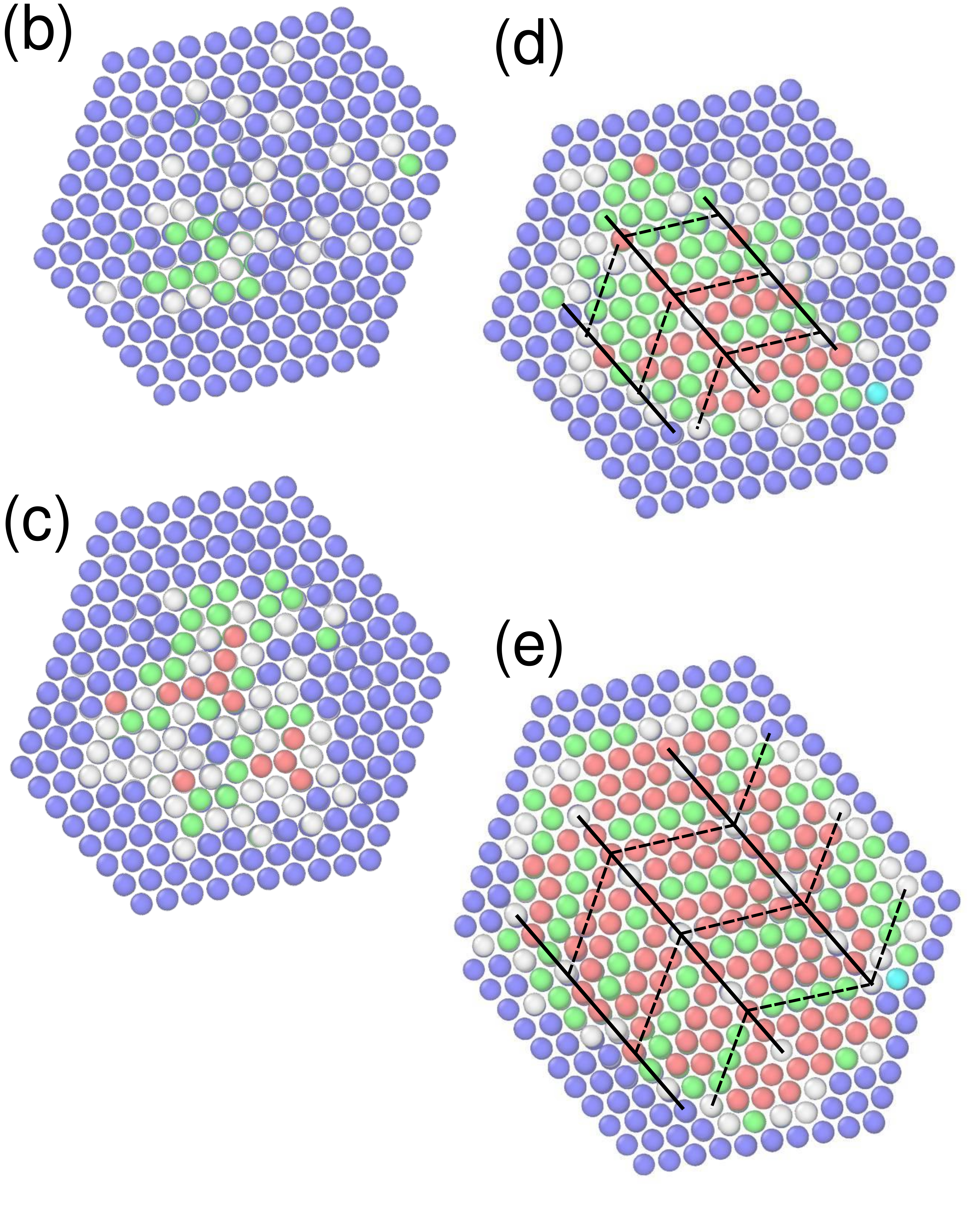}
  \caption{
    Precipitate structure at 200\,K from a VCSGC-MC/MD simulation during which the precipitate size was gradually reduced. The vertical dashed lines indicate the sizes of the precipitates shown in (b-e).
    (b) BCC precipitate after the transition (2,400 atoms),
    (c) precipitate in the intermediate region near the transition (2,900 atoms),
    (d) multiply-twinned 9R precipitate close to the transition (3,200 atoms),
    and
    (e) multiply-twinned 9R precipitate (5,550 atoms).
  }
  \label{fig:snapshots}
\end{figure}

For temperatures $\lesssim\,300\,\K$ the precipitate size at the transition can no longer support two twinned 9R domains. Under these conditions the transition occurs without hysteresis and with vanishing nucleation barrier as shown in \fig{fig:snapshots}. For sizes below 2,700 atoms the entire precipitate adopts a BCC structure. Between 2,700 and approximately 3,000 atoms the precipitate structure fluctuates rapidly between BCC and localized close-packed (CP) motifs while the average numbers of BCC and CP atoms vary approximately continuously with precipitate size. Eventually for larger precipitates multiply-twinned 9R structures are stabilized similar to the behavior at higher temperatures. Regardless of temperature and size 9R precipitates exhibit a core-shell structure comprising a CP core and a BCC wetting layer, the size of which varies between 1.5 and 3.5 atomic layers depending on temperature and size.

\subsection{Quenching simulations}
\label{sect:md_quenching}

\begin{figure}
  \centering
\includegraphics[width=0.95\linewidth]{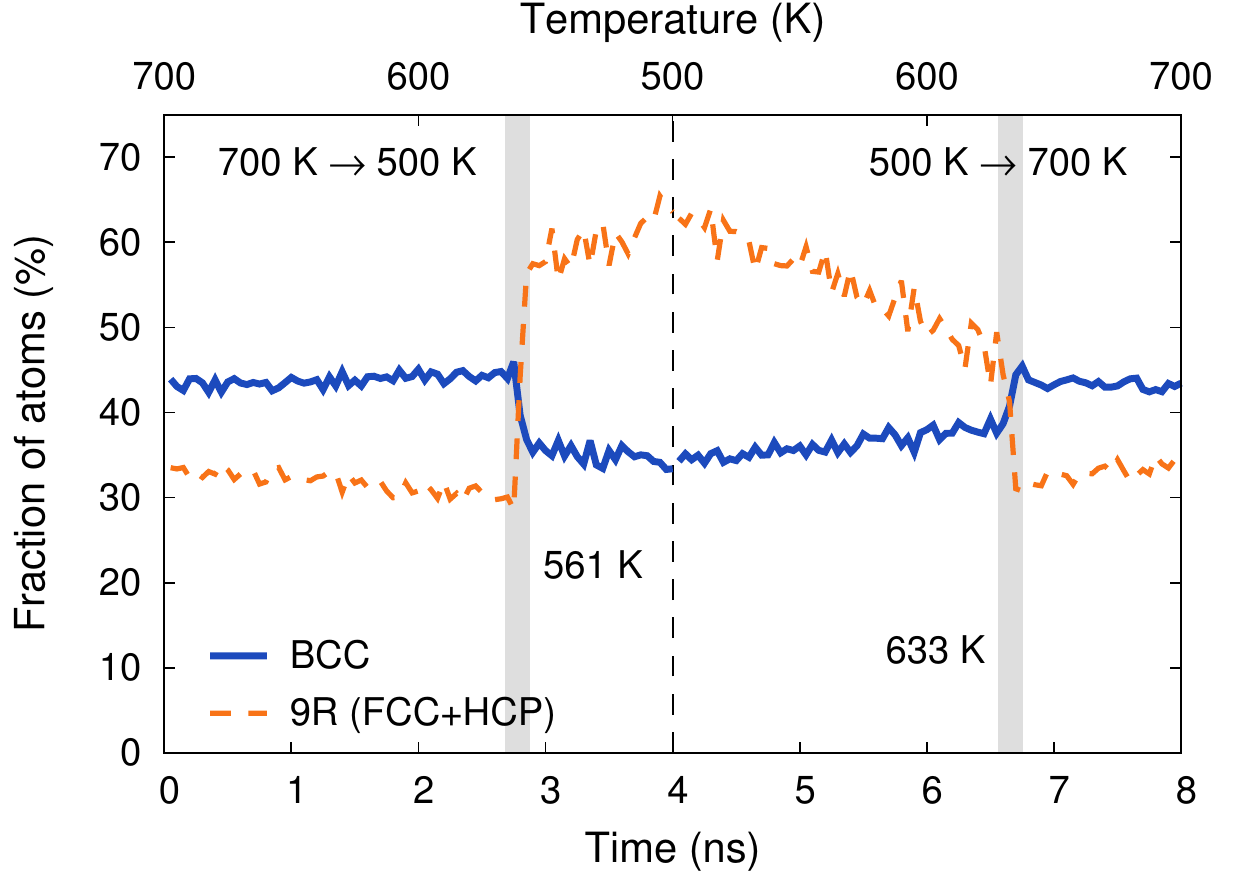}
  \caption{
    Fractions of atoms in Cu precipitate in different local environments during the course of a MD simulation. During the first 4\,ns of the simulation the temperature decreased linearly from 700 to 500\,K while during the following 4\,ns, the temperature was raised to its original value. The precipitate contained approximately 18,000 atoms. The transition temperatures obtained form the present simulation are plotted as open symbols in \fig{fig:temp_trans}.
  }
  \label{fig:md_quench}
\end{figure}

In \sect{sect:results_BCC_9R} it was shown that sampling the phase diagram in \fig{fig:temp_trans} horizontally (constant temperature) and vertically (constant concentration, approximately constant precipitate size) yields identical results when using VCSGC-MC/MD simulations. Since these simulations involve atom type swaps precipitates are allowed to change shape as they pass through the BCC--9R transition, which in reality requires diffusion of atoms near the Fe--Cu interface. Depending on the cooling rate in an experiment diffusion is not always possible and shape changes (or their suppression) could affect the phase transition.

To resolve this question a number of conventional MD simulations without atom type swaps were carried out, representing the limit in which diffusion is entirely suppressed (very fast quenching). BCC precipitates of varying size, which had been previously equilibrated about 100\,K above the expected transition temperature, were cooled down by 200\,K over a period of 4\,ns. Results from a typical simulation run are summarized in \fig{fig:md_quench}. It is observed that the transition is virtually instantaneous, providing further evidence for the martensitic character of the transformation. For the precipitate shown in \fig{fig:md_quench} the transformation to a multiply-twinned 9R structure occurs at a temperature of about 561\,K, which is notably lower than the transition temperature of 585\,K suggested by \fig{fig:temp_trans}.

To demonstrate that this slight temperature offset is not due to the large quenching rate, the temperature was ramped back to its original value. The reverse transition occurs again slightly {\em below} the temperature that one would expect based on \fig{fig:temp_trans} (633\,K {\it vs} 661\,K), which implies that the observed temperature difference cannot be explained by cooling/heating rate effects.
\footnote{
  The values for the fraction of atoms in BCC and FCC/HCP environments in \fig{fig:md_quench} are lower than in \fig{fig:trans_struct} because the Ackland-Jones analysis was carried out using instantaneous as opposed to time-averaged positions in order not to blur the temperature at which the structural transition occurs.
}

The temperature shift can rather be rationalized in terms of the different shapes of BCC and 9R precipitates. In the previous sections it was demonstrated that while BCC precipitates exhibit a spherical shape 9R precipitates resemble a prolate ellipsoid. The MD simulation described above started from a spherical BCC precipitate. Since shape changes were suppressed the spherical shape was also imposed on the 9R structures that formed during cooling. The spherical shape being less favorable for 9R precipitates increases the meta-stability range of BCC precipitates and causes a shift of {\em both} forward and backward BCC--9R transition to lower temperatures. As can be seen in \fig{fig:temp_trans} this behavior is observed consistently for precipitate sizes down to about 12,000 atoms. For even smaller precipitates the difference in shape between BCC and 9R precipitates becomes negligible and the shift in the transition temperature vanishes.

\section{Thermodynamics of BCC--9R transition}
\label{sect:results_thermodynamics}

\subsection{Size and temperature dependence of the transition}

\begin{figure}
  \centering
\includegraphics[scale=0.65]{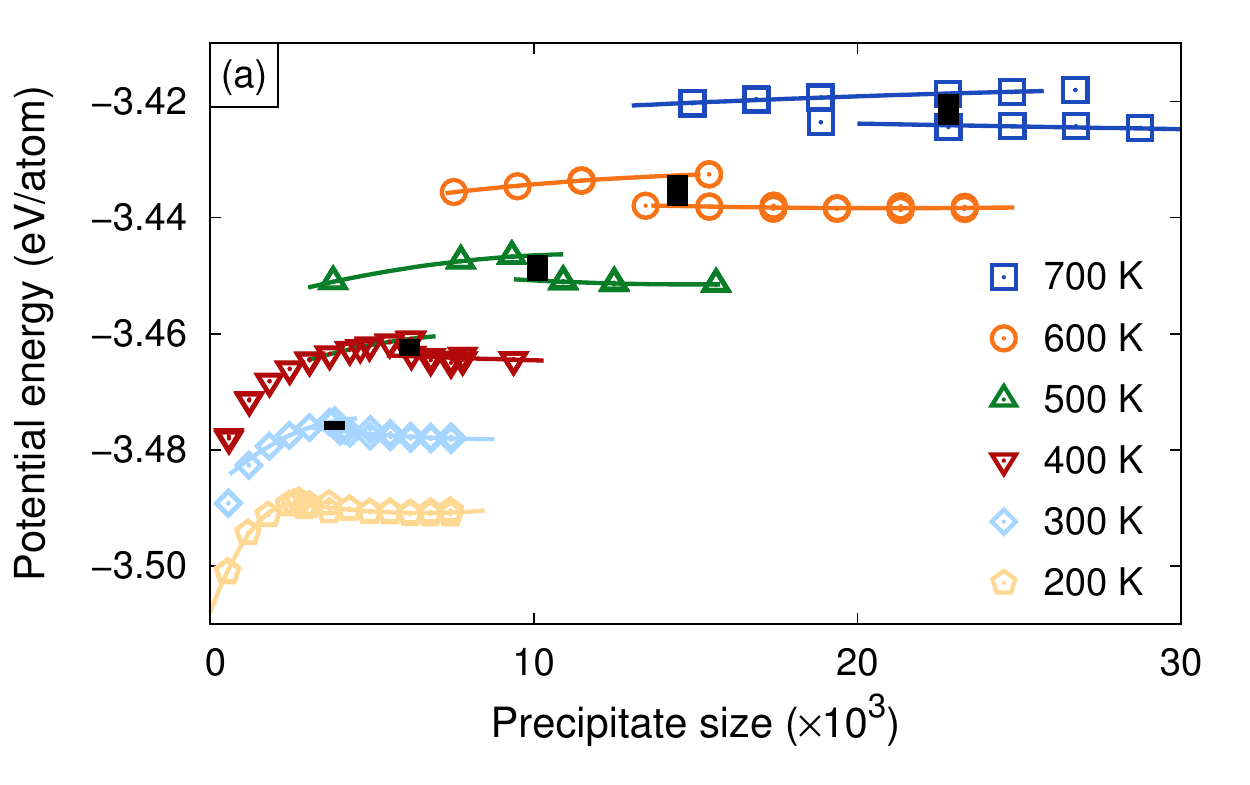}
\includegraphics[scale=0.65]{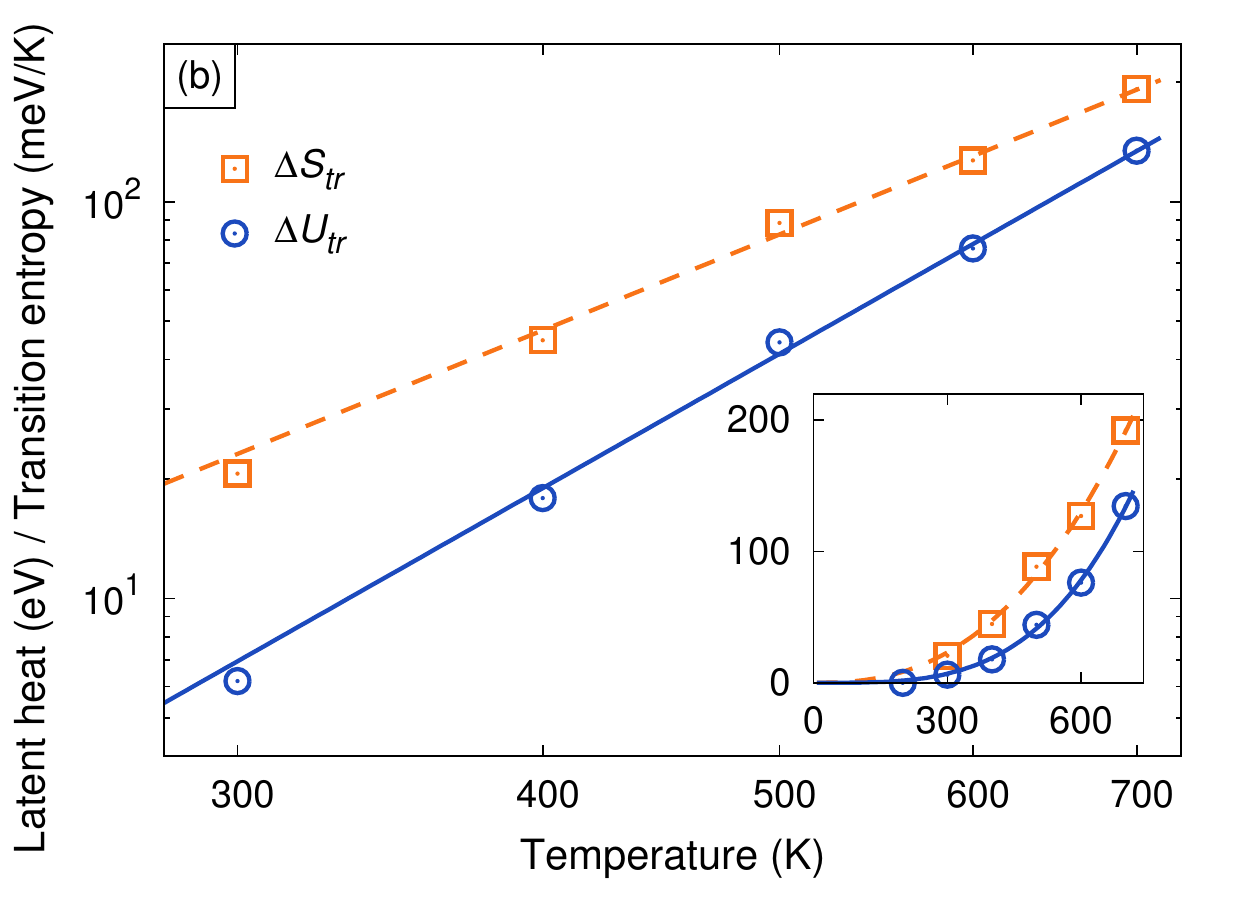}
  \caption{
    (a) Average potential energy of atoms in Cu precipitates as a function of precipitate size for different temperatures. The vertical black bars represent the heat of transformation, which is plotted in (b) as a function of temperature. Lines are guides to the eye.
  }
  \label{fig:epot_trans}
  \label{fig:latentheat}
\end{figure}

\begin{figure*}
  \centering
\includegraphics[scale=0.57]{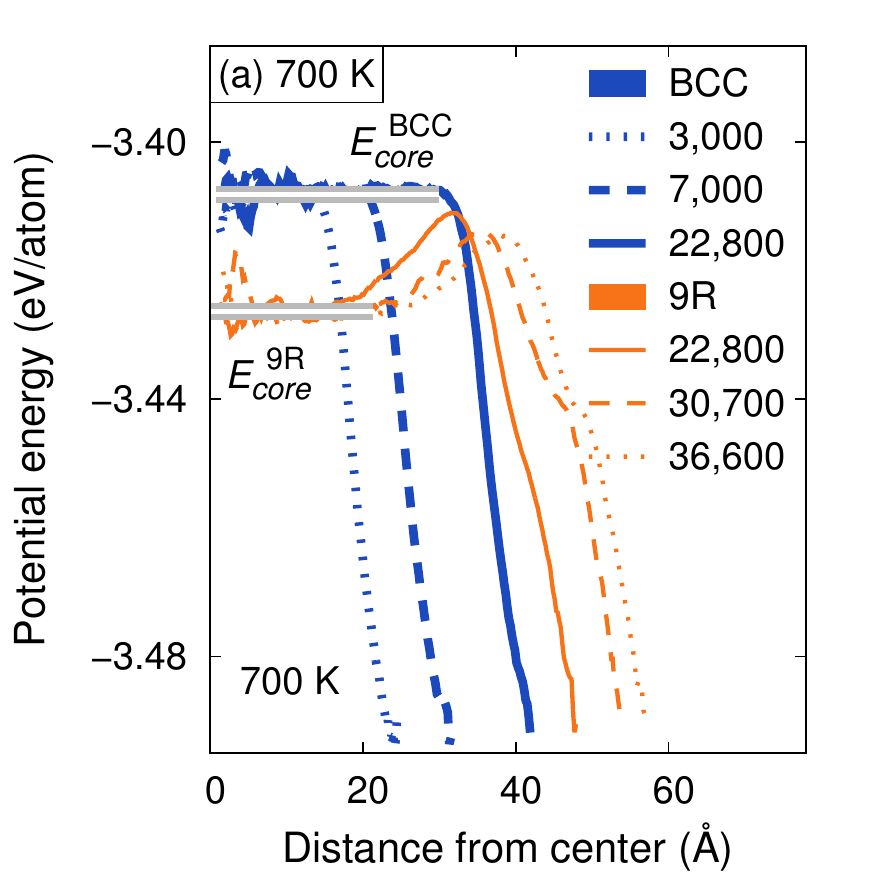}
  \hspace{-18pt}
\includegraphics[scale=0.57]{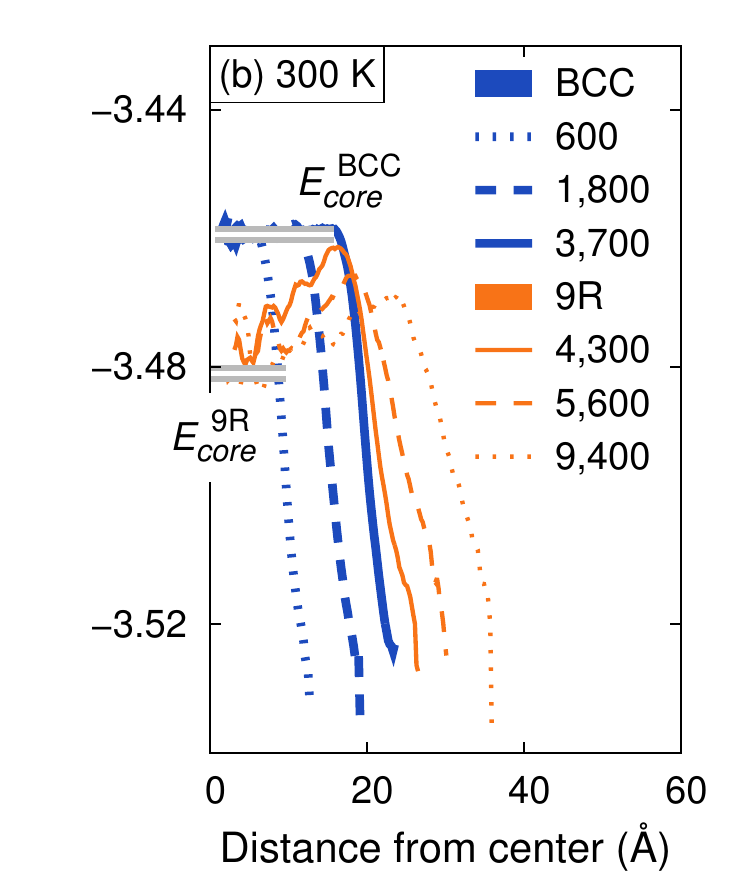}
  \hspace{-18pt}
\includegraphics[scale=0.57]{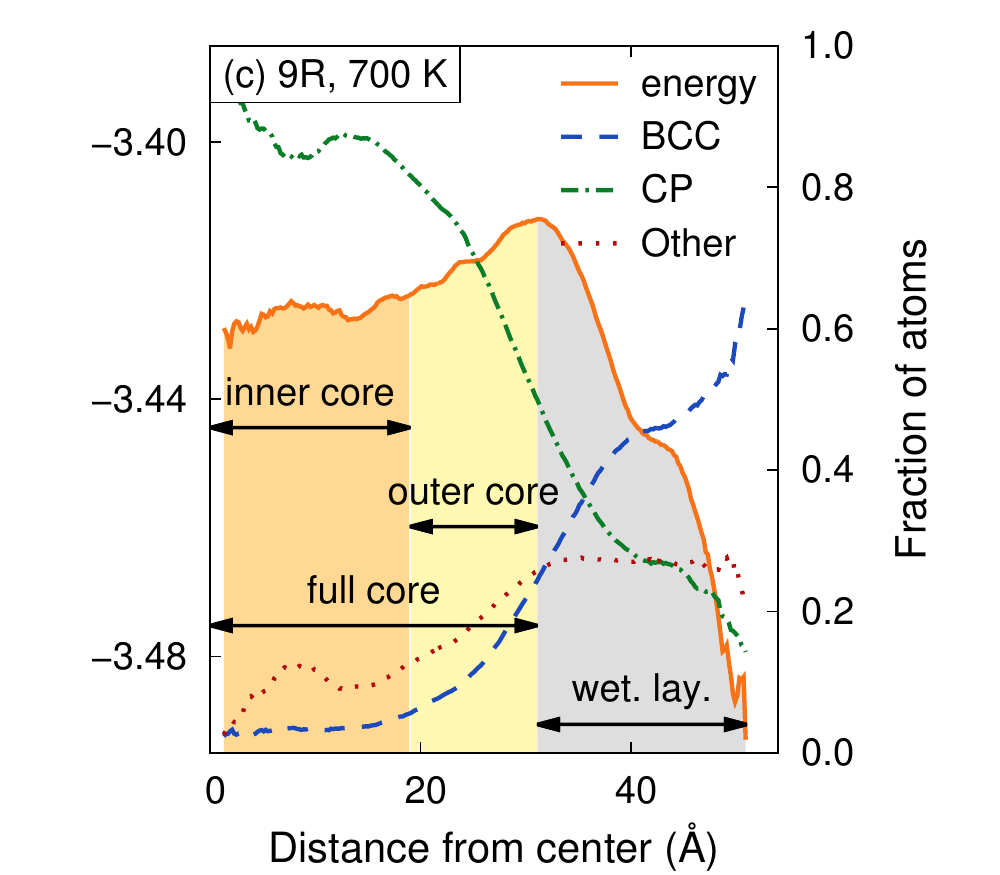}
\includegraphics[width=1.38in]{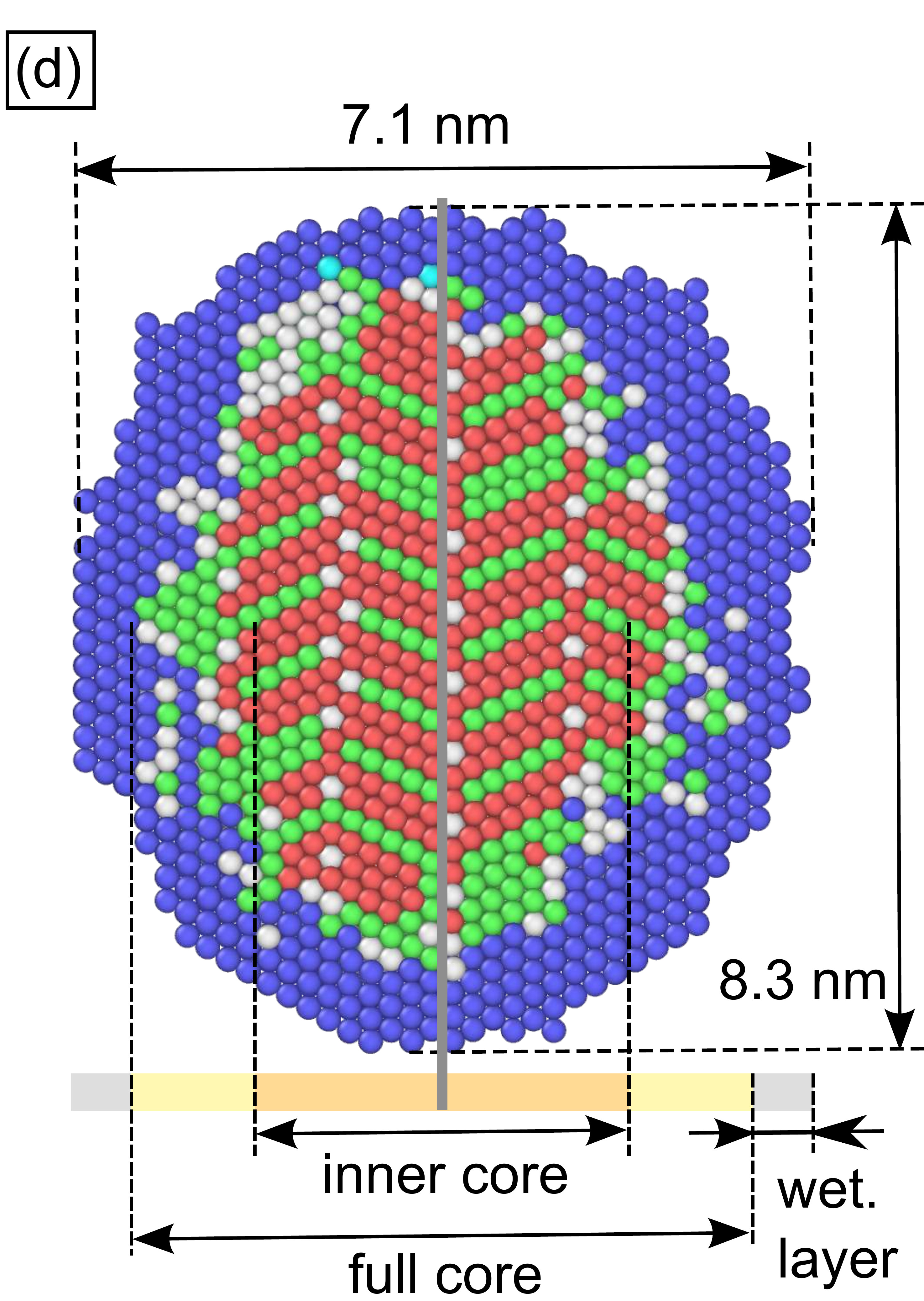}
  \caption{
    Potential energy per atom as a function of distance from precipitate center for BCC and 9R precipitates at (a) 700\,K (transition at about 22,800 atoms) and (b) 300\,K (transition at about 3,800 atoms).
    (c) Potential energy as well as fraction of number of atoms in different local environments for a precipitate containing 22,800 atoms at 700\,K.
    (d) Cross section of snapshot of precipitate shown in (c).
    The wetting layer region in (c) appears larger than in (d) due to the implicit projection of the elongated shape onto a sphere.
  }
  \label{fig:profiles}
  \label{fig:profile_structure}
\end{figure*}

The key result presented hitherto in this paper regards the phase diagram in \fig{fig:temp_trans}, which exhibits a strong increase in the size of the BCC-Cu precipitates with temperature before transition to the 9R phase occurs. This unusual size-temperature phase diagram can be explained by first noting that in bulk, the BCC-Cu phase is mechanically unstable. The unstable modes of the BCC-Cu phase are stabilized in small precipitates due to confinement and the restrictions imposed by the $\alpha$-Fe matrix. At finite temperatures, the entropy generated by these soft modes, then leads to stabilization of the BCC phase relative to the 9R phase.   

Even more interesting is the significant change that is observed in the character of the phase transformations as a function of temperature. At high temperatures, the BCC--9R transitions exhibit strong first-order character and prominent hysteresis. Upon cooling the discontinuities become less pronounced and the transitions occur at ever smaller cluster sizes. Below 300\,K the hysteresis vanishes while the latent heat diminishes.

To provide a more quantitative understanding of the character of the transition we have analyzed the size and temperature dependence of the potential energy of the atoms in the Cu precipitate
\footnote{
  The analysis of the potential energy has been carried out both by considering all atoms in the system (Fe matrix as well as Cu precipitate) and the Cu precipitate only with practically identical results. For the sake of clarity here we only show data for atoms in precipitates.
}
as shown in \fig{fig:epot_trans}(a). At higher temperatures the data exhibits a pronounced discontinuity as expected for a first-order transition. As the temperature is lowered the precipitate size at which the transition is observed decreases, and so does the discontinuity in the potential energy.  It practically vanishes below 300\,K, in the same temperature range in which the hysteresis disappears. This aspect is illustrated in \fig{fig:latentheat}(b), which shows strong variation of the latent heat with temperature. In the remainder of this section we will focus on resolving the origin of the unexpectedly large temperature dependence of the structural transformation of the Cu precipitates.

\subsection{Microscopic analysis}
\label{sect:results_thermodynamics_profiles}

Figure~\ref{fig:profiles} shows the potential energy as a function of distance from the precipitate center for both BCC and 9R precipitates at two different temperatures, and illustrates the correlation between structure and potential energy profile. BCC precipitates exhibit a clearly defined core region, in which the potential energy is constant. For a given temperature this energy, which can be interpreted as the value equivalent to bulk BCC Cu, is size independent.
\footnote{
  This statement excludes very small precipitates ($N\lesssim\,500\,\text{atoms}$), for which the wetting layer region comprises the entire precipitate. It is for the latter reason that in \fig{fig:epot_trans}(a) the potential energy of small precipitates decreases rapidly with size.
}
Outside the core radius the potential energy drops toward the Fe--Cu interface. This represents the wetting layer between BCC Cu core and BCC Fe matrix.

The profiles for larger 9R precipitates [\fig{fig:profiles}(a)] also reveal a plateau in the core region, which is, however, smaller in terms of radius than for BCC precipitates of the same size. Outside this inner core region the potential energy increases up to a maximum value [``full core'' region in \fig{fig:profile_structure}(c,d)] before it decays toward the Fe--Cu interface in a similar fashion as for the BCC precipitates (``wetting layer'' region). Note that the shoulder that appears for example in the wetting layer region of the energy profile in \fig{fig:profile_structure}(c) is due to the projection of an elongated ellipsoidal precipitate onto a sphere (compare \fig{fig:trans_shape}). For the same reason the wetting layer seemingly has a larger spatial extent in \fig{fig:profile_structure}(c) than in \fig{fig:profile_structure}(d).

As shown in \fig{fig:profile_structure} the different regions in the energy profile can be correlated with different atomic structures. The inner core region corresponds to the close-packed core of the precipitate representing bulk 9R, whereas the outermost region comprises BCC atoms and acts as a wetting layer between Cu and Fe. The intermediate (outer core) region is characterized by a continuous variation in the {\em average} number of BCC and close-packed atoms as well as the potential energy. This region is required to connect the BCC and close-packed lattices crystallographically, and corresponds to the outermost layers of the 9R region of the precipitate for which the misorientation angle is reduced from its maximum value as shown in \fig{fig:angle_trans}(a). It thus corresponds to a close-packed region under shear strain. As will be discussed below this region is crucial for understanding the strong reduction in the latent heat with decreasing temperature (and transition size).

\begin{figure}
  \centering
\includegraphics[scale=0.62]{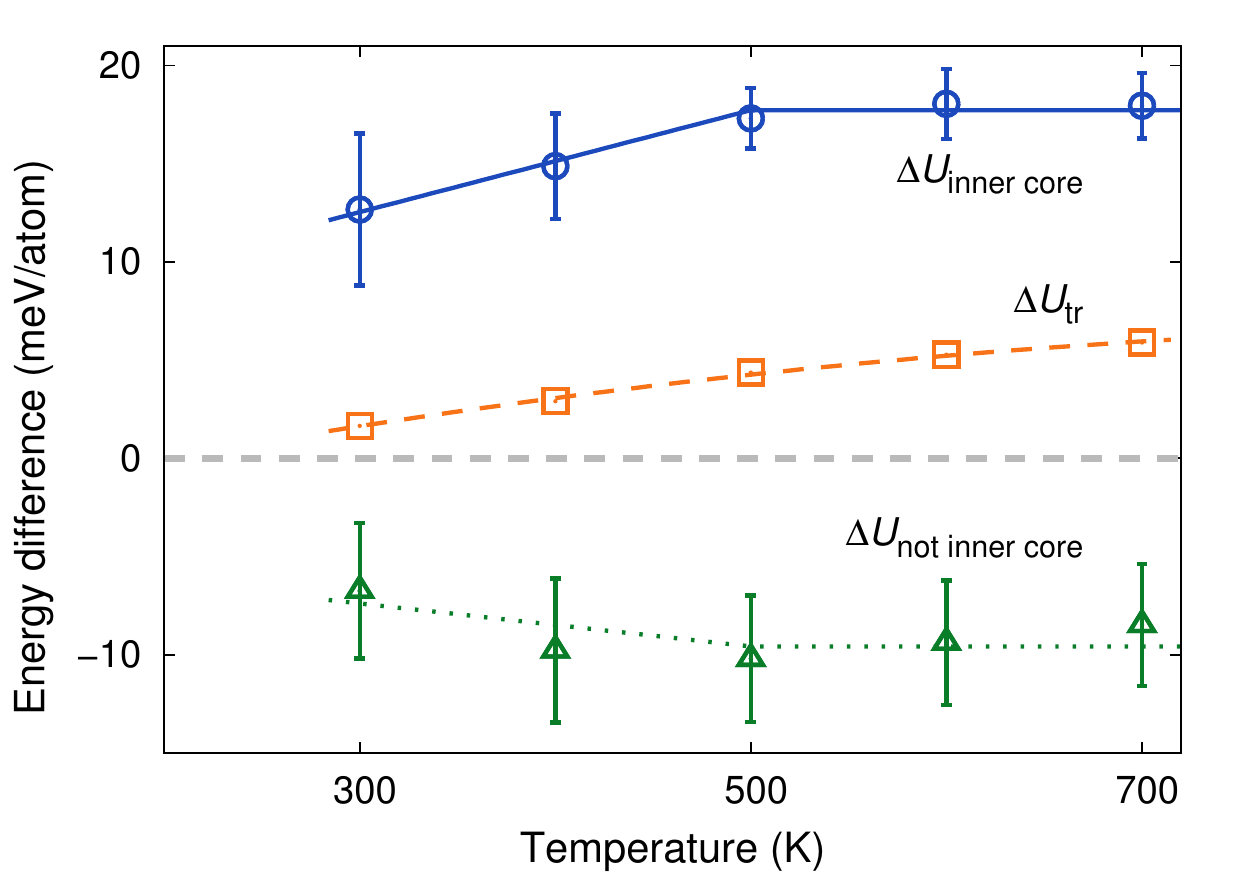}
  \caption{
    Total latent heat ($\Delta U_{tr}$) obtained from analysis of average potential energy (see \fig{fig:latentheat}) and region-decomposed energy contributions ($\Delta U_\text{core}$, $\Delta U_\text{not inner core}$) obtained from analysis of energy profiles of the type shown in \fig{fig:profiles}.
  }
  \label{fig:ecore_vs_temp}
\end{figure}

For large precipitates ($\gtrsim\!7,000\,\text{atoms}$) the inner core (``bulk'') potential energy of 9R precipitates is practically size independent as exemplified in \fig{fig:profiles}(a). For smaller precipitates, however, the inner core region can no longer fully relax since the confinement constrains the misorientation angle, which raises the energy compared to larger 9R precipitates [\fig{fig:profiles}(b)].

We now discuss in detail the core and wetting layer contributions to the latent heat for different temperatures as shown \fig{fig:ecore_vs_temp}. Both contributions approach asymptotic values, which for the core amounts to 20\,meV/atom. This is about half the value of 40\,meV/atom that is found for the energy difference between the ideal BCC and 9R structures at zero Kelvin. The lower value obtained from direct analysis of the precipitate core energies can be attributed to the high planar fault density in actual 9R precipitates [twin planes as well as stacking faults, see Figs.~\ref{fig:viz_prec} and \ref{fig:profile_structure}(d)].

While the core potential energy difference between BCC and 9R precipitates is positive, \fig{fig:ecore_vs_temp} shows that the wetting layer energy difference has the opposite sign. This feature, in addition to the wetting layer increasing in size in smaller precipitates, is responsible for the drastic reduction in total latent heat as temperature is lowered (see Figs.~\ref{fig:latentheat} and \ref{fig:ecore_vs_temp}). This is because at lower temperatures the transition occurs at smaller precipitate sizes, where the number of Cu particles belonging to core and wetting layer begin to equalize, leading to mutual cancellation of their contributions to the latent heat. 

\begin{figure}
  \centering
\includegraphics[width=0.95\linewidth]{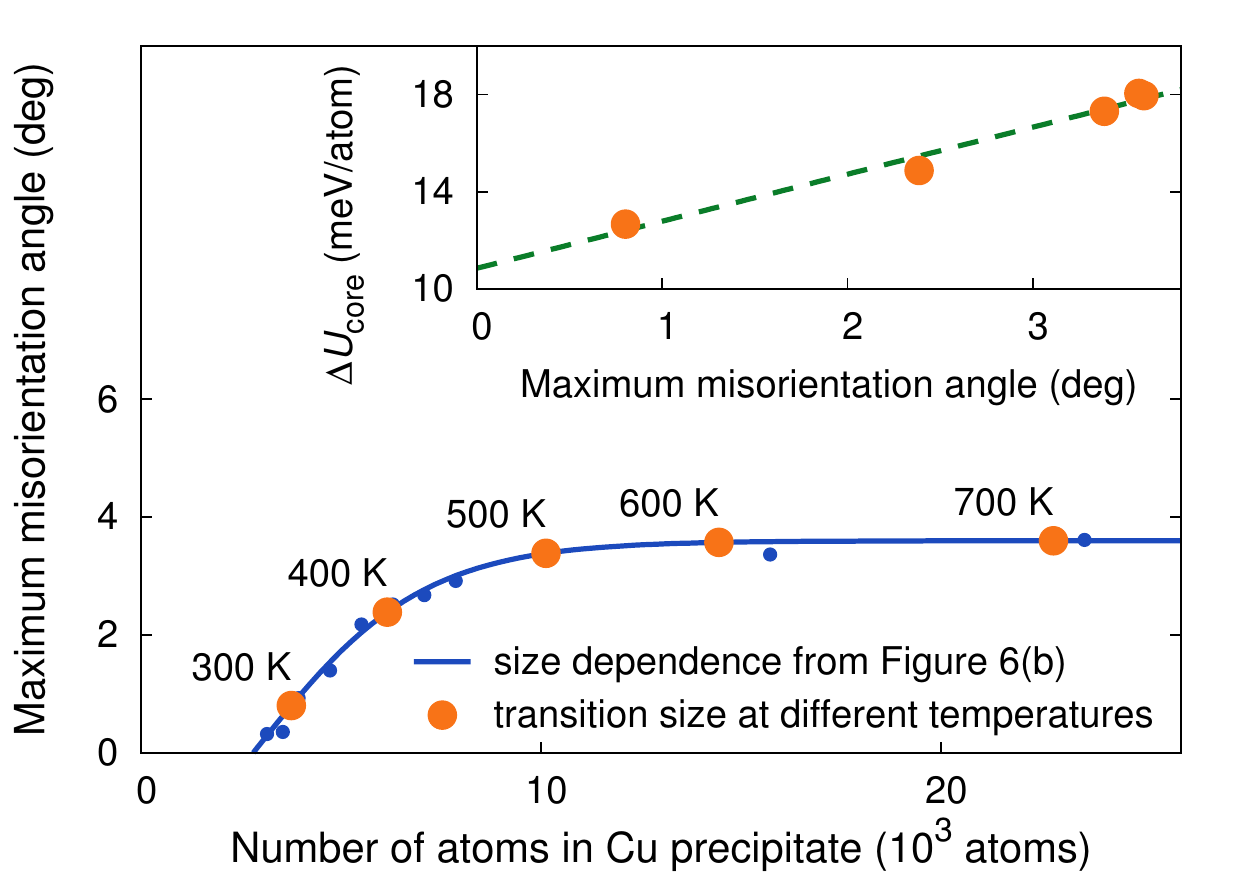}
  \caption{
    Maximum misorientation angle as a function of precipitate size. The large filled circles indicate the maximum orientation angle observed at the transition size for different temperatures. The inset shows that the core energy differences $\Delta U_\text{core}$ from \fig{fig:ecore_vs_temp} scale approximately linearly with the maximum misorientation angle.
  }
  \label{fig:misorientation}
\end{figure}

We conclude the description of energy profiles by providing further insight on the effect of strain on the core energetics, which resides in the misorientation angle data presented in \fig{fig:angle_trans} and \sect{sect:results_9R}. Figure~\ref{fig:angle_trans}(a) demonstrates that the misorientation angle is not constant across the precipitate but rather increases continuously from zero near the interface to its maximum value in the core region of the precipitate. This observation explains the gradual change in the potential energy profile in the outer core region in \fig{fig:profile_structure}. As already mentioned in \sect{sect:results_9R} the maximum misorientation angle is practically temperature independent and levels off at approximately 3.6\deg\ in the limit of large precipitates. It is fundamentally determined by the 9R twin boundary energetics. Using the data in \fig{fig:angle_trans}(b) one can demonstrate a simple correlation between misorientation angle and the core contribution to the latent heat. Figure~\ref{fig:misorientation} shows a fit to the data from \fig{fig:angle_trans}(b) and indicates the maximum misorientation angles that are observed at transition for different temperatures. While the data points above 500\,K all yield approximately the same maximum misorientation angle, the 300 and 400\,K data points correspond to considerably smaller values. As shown in the inset the core potential energy difference exhibits a simple linear dependence on the maximum misorientation angle and is accordingly correlated with strain.

As can be seen in \fig{fig:ecore_vs_temp} the confinement strain has a stronger effect on the energy difference in the core region than in the wetting layer. In fact below 300\,K when the transition occurs without hysteresis, the misorientation angle also becomes vanishingly small. According to the fit in \fig{fig:misorientation} the latter should become zero at a size of about 2,800 atoms, which correlates very well with the size range between 2,700 and 3,000 atoms, for which one can no longer observe stable 9R twins in simulations at 200\,K (see \fig{fig:snapshots} and \sect{sect:results_transmech}).

\section{Dislocation--precipitate interaction}
\label{sect:results_dislocations}

Ultimately, Cu precipitation in ferritic alloys leads to hardening by impeding dislocation motion. Hardening is a macroscopic quantity that takes into account multiple dislocation-precipitate interactions over a statistically significant volume. However, the elementary hardening mechanism in our case is dislocation bowing around a precipitate, which occurs at the atomic scale and is amenable to study by MD simulations. In fact, recent simulations show that the dislocation core, which represents the very atomic essence of a dislocation, may suffer structural transformations when near or inside a precipitate. \cite{HarBac02, CheKioGho09}

Similarly, in the presence of a stress field, such as that of a dislocation, grain boundary, etc., Cu precipitates may suffer internal changes. Indeed, Cu BCC precipitates have been observed to locally transform to a close-packed structure when subjected to the stress field of a screw dislocation. \cite{ShiChoKwo07, ShiKimJun08, BacOse09}

The incremental increase in shear stress $\tau_p$ due to Cu precipitates can be estimated from a dispersed barrier hardening (DBH) model \cite{Luc93} with knowledge of size, number density and composition of precipitates,
\footnote{
  It is worth noting that, although the DBH model is not successful in explaining many aspects of deformed irradiated materials (see, e.g., Ref.~\onlinecite{SinForTri97}), it provides a simple framework to show the connection between MD simulations of dislocation-precipitate interaction and continuum hardening laws.
}
\begin{align}
  \tau_p = \frac{\alpha\mu b}{L}
  \label{eq:dbh1}
\end{align}
where $\mu$ is the matrix (BCC Fe) shear modulus, $b$ is the Burgers vector, $L$ is the average barrier spacing, and $\alpha$ is the strengthening coefficient. Estimates of $\alpha$ can be obtained by recourse to line tension expressions of the DBH, such as the Fleischer-Friedel formula, \cite{Fri64}
\begin{align}
  \tau_p = \frac{\mu b}{L}\cos{\left(\frac{\phi_c}{2}\right)^{\frac{3}{2}}}
  \label{eq:dbh2} 
\end{align}
where $\phi_c$ is the critical cutting angle after which the dislocation is able to traverse the obstacle. Measurements of $\phi_c$ can be obtained via, e.g., {\it in-situ} TEM observations of dislocation-irradiation obstacle interactions. \cite{RobBeauFad05} MD can provide direct estimates of $\tau_p$, although it suffers from length and time scale limitations that result in unphysically high strain rates. \cite{NedKizSch00, Mar02, BacOse04, BacOse09}

\begin{figure*}
  \centering
  \includegraphics[width=0.95\linewidth]{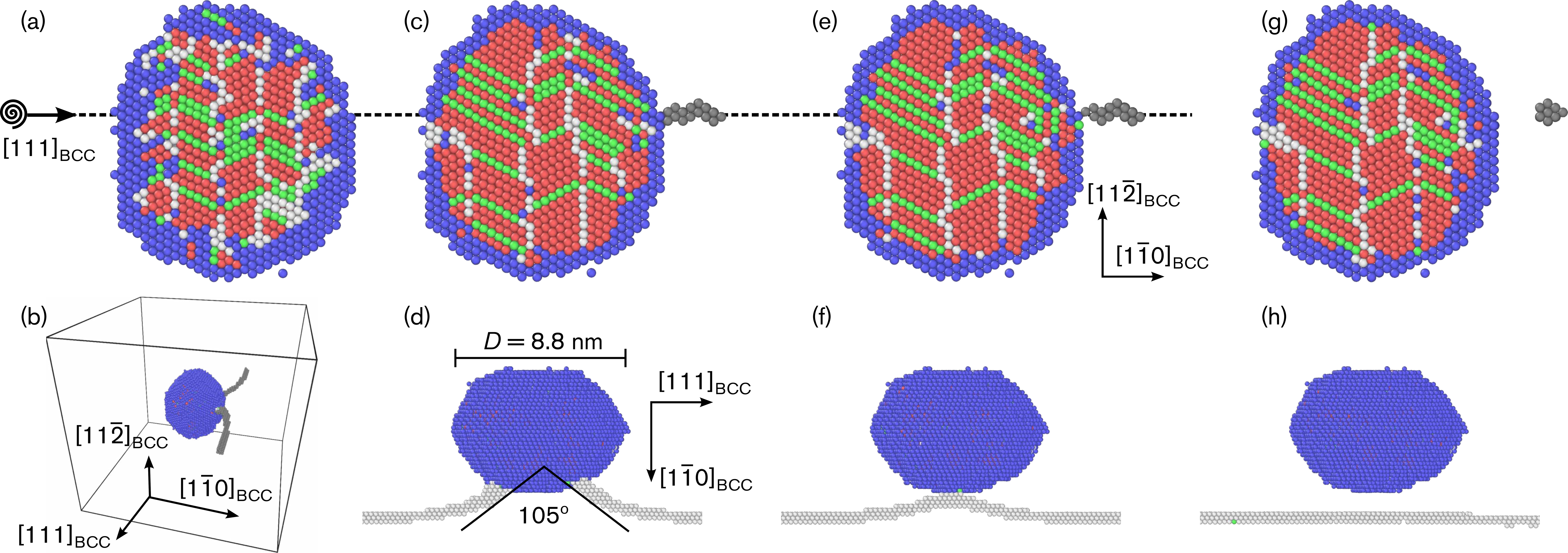}
  \caption{
    Illustration of screw dislocation (line vector parallel to $[111]_\text{BCC}$ and $[\bar{1}10]_\text{9R}$) interacting with a 9R precipitates containing 17,400 atoms ($D=8.8\,\nm$, compare \tab{tab:data}). The screw dislocation moves along $[1\bar{1}0]_\text{BCC}$ as indicated by the dashed line.
    (a) Initial precipitate. (b--d) Dislocation inside precipitate just prior to detachment. (e-h) Dislocation as it exits precipitate.
    The magnitude of the exit angle expected based on \eq{eq:dbh2} is indicated in (d).
  }
  \label{fig:exitangle}
\end{figure*}

The crystallographic orientation of the precipitate adds yet another dimension to the already multidimensional character of the dislocation-precipitate interactions. Most of the published works on dislocation-precipitate interactions in Fe--Cu alloys consider small coherent BCC precipitates, \cite{BacOse04, KohKizSch05, BacOse09} both in order to study incipient hardening at the beginning of Cu precipitation and due to the absence of high-fidelity atomistic models of 9R and FCC Cu precipitates. Our work effectively eliminates this limitation for the first time. Here, we study screw dislocation interactions with several precipitates obtained by quenching from $600$ to $10\,\K$ including both BCC and 9R configurations, and compute $\tau_p$ as a function of $L$.
Although statistically speaking the average barrier spacing in \eq{eq:dbh1} $L=\left(2Nr_\Cu\right)^{-1/2}$ is taken from the known number density $N=\left(84a\times30b\times52c\right)^{-1}$ and diameter $D$ of the precipitates, here we simply take $L=84a-D$ for consistency with the MD setup. We then measure the critical shear stress required to overcome the obstacle, the results for which are given as $\tau_p$--$L$ data pairs in Table~\ref{tab:data}. Snapshots from a typical simulation illustrating the process of a dislocation cutting the precipitate are shown in \fig{fig:exitangle}.

\begin{table}
  \centering
  \caption{
    Calculated precipitate strength for precipitates of different sizes where $L$ is the dislocation line length and $D$ the precipitate diameter parallel to the dislocation line. The orientation column indicates the vector in the frame of reference of the precipitate that aligns with the screw direction in the matrix. The last precipitates are nominally equal in size but the orientation with respect to the dislocation line differs. As a result of the elongated shape of the 9R precipitate (compare Figs.~\ref{fig:viz_prec} and \ref{fig:trans_shape}) the latter has a shorter effective diameter projected onto the dislocation line.
  }
  \label{tab:data}
  \begin{tabular}{*{6}{c}}
    \hline\hline
    Size & $L$ (nm) & $D$ (nm) & $\tau_p$ (MPa) &  Structure & Orientation \\
    \hline
    600    & 16.4 & 4.4 &  320 & BCC & $[111]_\BCC$ \\
    7,500  & 15.4 & 5.4 &  450 & 9R  & $[\bar{1}10]_{9\R}$ \\
    10,900 & 13.6 & 7.2 &  675 & 9R  & $[\bar{1}10]_{9\R}$ \\
    17,400 & 12.0 & 8.8 & 1150 & 9R  & $[\bar{1}10]_{9\R}$ \\[6pt]
    17,400 & 10.0 & 6.8 & 2250 & 9R  & $[11\bar{4}]_{9\R}$ \\
    \hline\hline
  \end{tabular}
\end{table}

Data points corresponding to the $[\bar{1}10]_{9\R}$ orientation can be fit to an equation of the form $\tau_p=AL^{-1}$, which yields $A=8.42\times10^{-6}$ GPa\,nm. Using values
\footnote{
  Values given correspond to those obtained after rotating the nominal shear modulus of 86\,GPa to the slip geometry used here.
}
of $b=0.25\,\text{nm}$ and $\mu=71.5\,\text{GPa}$, we obtain $\alpha=0.47$ for the strengthening coefficient in \eq{eq:dbh1}, which using \eq{eq:dbh2} results in an effective exit angle of $105^{\circ}$. This is in good agreement with the direct analysis of the simulations as shown in \fig{fig:exitangle}(d).

Regarding the interaction mechanism, in the previous sections it has been demonstrated that the $\langle111\rangle$ directions of the Fe matrix are aligned with the $\langle110\rangle$ directions of 9R precipitates, which are close-packed. This implies that when a $\small{\frac{1}{2}}\langle111\rangle$ screw dislocation line is co-linear with this close-packed direction, the screw character of the dislocation can be conserved within the precipitate. Then, the resistance exerted by the precipitate on the dislocation originates from the energy cost associated with the core energy of the 9R dislocation as well as from the herring-bone slip plane structure encountered inside the precipitate. This structure must be negotiated via multiple cross-slip steps that may hinder dislocation motion as well.

When the principal $\langle110\rangle_{9\R}$ axis is not aligned with the screw direction, the dislocation must nucleate a set of interface dislocations to conserve the Burgers vector. In that case, the energetics associated with these nucleation episodes result in considerably higher hardening. To exemplify this effect, we have studied one such interaction corresponding to the orientation $[11\bar{4}]_{9\R}$, which results in a critical stress that is two times higher than for the first orientation. The calculated $\tau_p$ value is also given in \tab{tab:data} and included for comparison in \fig{fig:stress}.

\section{Discussion}
\label{sect:discussion}

\subsection{Thermodynamic implications}
\label{sect:discussion_thermodynamics}

The transition diagram in \fig{fig:temp_trans} represents the first key result of the present study. Most importantly it demonstrates that the precipitate size at which the transition occurs varies strongly with temperature, changing from 2,700 atoms (equivalent to a precipitate diameter of $\sim\!4\,\text{nm}$) at 200\,K to 22,800 atoms ($\sim\!8\,\text{nm}$) at 700\,K. This strong size dependence is a direct result of the soft vibrational modes of the BCC Cu phase, which are responsible for this phase having a larger vibrational entropy than 9R Cu. Hence increasing temperature leads to stabilization of larger BCC-Cu precipitates. As will be discussed in \sect{sect:discussion_experiment} below the temperature dependence of the transition size is in very good agreement with the interpretation of experimental data obtained by Monzen and co-workers \cite{MonJenSut00}.

The second important finding pertains to the order of the transition. At higher temperatures it exhibits pronounced hysteresis and a finite latent heat (6\,meV/atom at 700\,K, see \fig{fig:ecore_vs_temp}). With decreasing temperature the latent heat diminishes and the nucleation barrier vanishes at approximately 300\,K, whereupon the transition occurs without hysteresis. The disappearance of the nucleation barrier is accompanied by characteristic changes in the misorientation angle. For precipitates below approximately 6,000 atoms (5\,nm) it quickly decreases and reaches zero at a size of 2,800 atoms, which correlates with the critical size observed at 200\,K. The strong size dependence of the misorientation angle is the result of the crystallographic boundary conditions along the interface between close-packed (inner core in \fig{fig:profiles}) and BCC Cu (wetting layer) regions, which are manifested in the presence of a layer of close-packed Cu atoms under shear strain (outer core).

The ingredients that underlie the strong size dependence and the change in transition character are:
({\em i}) two crystallographically mismatched boundary phases
and
({\em ii}) the minority component is dynamically unstable or at least vibrationally soft in the lattice structure of the matrix phase.
These conditions are very general and can be found in a large number of systems. The features described in the present work should therefore also be observable in other systems.

Vacancies are known to play an important role in the formation of Cu precipitates in iron. \cite{Ode83, SoiBarMar96, DomBecVan99, VinBecPar08} They have also been discussed in the context of the BCC--9R transformation (see e.g., Ref.~\onlinecite{OseSer97, BlaAck01}). While {\em thermodynamically} vacancies are not needed for the transition they are likely to play a crucial role for the kinetics of the transformation and the loss of coherency between BCC matrix and 9R precipitates (see \sect{sect:discussion_experiment}).

\subsection{Implications for interpretation of experiments}
\label{sect:discussion_experiment}

Most TEM studies (see e.g., Refs.~\onlinecite{OthJenSmi91, OthJenSmi94, LeeKimKim07}) were carried out at room temperature on samples that were quenched from temperatures above 750\,K to room temperature. The smallest 9R precipitates observed in these studies were typically just slightly larger than 4\,nm in excellent agreement with the critical size at about 300\,K obtained in the present simulations.

In the present context comparison with the experimental study by Monzen and co-workers \cite{MonJenSut00} is particularly interesting. Motivated by some earlier results by Habibi-Bajguirani and Jenkins \cite{HabJen96}, these authors carefully investigated the size dependence of the angle $\phi$ enclosed by $\{001\}_\text{9R}$ planes in adjacent twin segments in 9R precipitates as a function of annealing temperature.

They observed that precipitates annealed for one hour at 673\,K exhibit a value  of $\phi\approx 128\deg$ up to a size of about 9\,nm. For larger precipitates the angle drops sharply to $\phi\approx 122\deg$. The size at which this transition takes place is strongly dependent on the annealing temperature, with lower temperatures leading to the transition occurring at smaller sizes. Monzen {\it et al.} interpreted this finding as an indication for a temperature-dependent transition between BCC and 9R structures strongly reminiscent of the present findings. The experimental data points are included in \fig{fig:temp_trans} and in fact display a remarkable agreement with the BCC--9R transition line obtained from simulation.To appreciate the interpretation by Monzen {\it et al.} and the connection to the present simulations let us revisit the experimental observations in more detail. As discussed extensively in \sect{sect:results_9R} $\{001\}_\text{9R}$ planes are approximately parallel to ${110}_\text{BCC}$ planes. The deviation from perfect alignment is measured by the misorientation angle (see \fig{fig:schematic_9R_displ}). The value of $\phi=128\deg$ observed by Monzen and co-workers for smaller precipitates is equivalent to a misorientation angle of about $\theta=4\deg$, which agrees both with earlier measurements on quenched precipitates \cite{OthJenSmi91, OthJenSmi94, LeeKimKim07} and our simulations. The value of $\phi=122\deg$ obtained for larger precipitates on the other hand corresponds to a misorientation angle $\theta$ of only 1\deg. The latter is the result of ---presumably diffusion mediated--- reshuffling at the interface, by which the strain in the 9R precipitates is reduced. \cite{MonJenSut00} It is likely that vacancies play a crucial role in this process (compare Refs.~\onlinecite{OseSer97, BlaAck01}).

A precipitate that adopts the 9R structure at the annealing temperature will be able to rearrange its interface during the extended annealing period and thereby reduce its misorientation angle. In contrast a precipitate that adopts the BCC structure under the annealing conditions will rapidly pass through the BCC--9R transition during quenching with insufficient time and activation to transform its interface. As a result it will exhibit a misorientation angle of about 4\deg\ as dictated by the interface alignment with the Fe BCC matrix and the Cu BCC wetting layer. This distinction enables the indirect experimental detection of the BCC--9R transition by TEM observation at room temperature.

In our simulations we directly observe the diffusion-less transition that takes place during rapid quenching (or heating). The time scale of our simulations is too short to allow for a diffusional transformation of the interface and as a result the simulated precipitates exhibit maximum misorientation angles of up to 4\deg. It is important to stress that the experimental observation is only possible {\it because of diffusion} along the interface since otherwise 9R precipitates that undergo a transition during quenching would be indistinguishable from precipitates that adopt the 9R structure already during annealing. In this sense experiment and simulation provide complimentary views, and while the direct simulation of the BCC--9R transition confounds the interpretation of the experimental observation, the good agreement with experiment verifies the simulation.

It remains to point out that a diffusional interface transformation is unlikely to be of importance for the low temperature behavior described in \sect{sect:results_thermodynamics} not only because diffusion is hampered at such low temperatures but ---more importantly--- because for the relevant precipitate sizes the misorientation angle is already strongly reduced due to confinement (compare \fig{fig:angle_trans}).

To conclude this section a few additional comments regarding the structure and shape of Cu precipitates are in order. In Sects.~\ref{sect:results_BCC} and \ref{sect:results_9R} it was found that BCC precipitates adopt a spherical equilibrium shape whereas 9R precipitates resemble an elongated ellipsoid. Conversely, in TEM investigations smaller 9R precipitates appear often to be characterized as circular (in 2-D projection), whereas larger 9R precipitates show an elongated shape. This observation is readily understood on the basis of the present data as smaller circular 9R precipitates seen in TEM investigations actually correspond to (spherical) BCC precipitates at annealing temperatures. During rapid quenching to room temperature these precipitates transform into the 9R structure. Shape change, however, is suppressed since it once again would require diffusion. In contrast larger precipitates adopt the 9R structure also under annealing conditions and remain 9R during cooling. This situation resembles the MD simulation shown in \fig{fig:md_quench}.

The simulated precipitates exhibit some features that are either impossible or very difficult to detect experimentally. For example \fig{fig:viz_prec}(b) reveals that multiply-twinned 9R Cu precipitates have a BCC Cu shell that typically comprises about two to three atomic layers and serves as a wetting layer between the close-packed Cu core and the BCC Fe matrix. There is thus no direct 9R Cu and BCC Fe interface, which is of uttermost importance for the thermodynamics of the transition as discussed in detail above. The existence of a wetting layer has some additional implications including the following:
({\it i}) 
Experimentally, the diameter of Cu precipitates is typically measured using contrast produced by different lattice structures. As this probe is insensitive to the BCC Cu shell it leads to a slight underestimation of precipitate size.
({\it ii}) A dislocation interacting with a 9R Cu precipitate will not see a sharp BCC Fe|9R Cu interface but instead pass pass through three distinct zones (BCC Fe, BCC Cu, and 9R Cu).

\subsection{Dislocations}
\label{sect:discussion_dislocations}

\begin{figure}
  \centering
\includegraphics[width=0.95\linewidth]{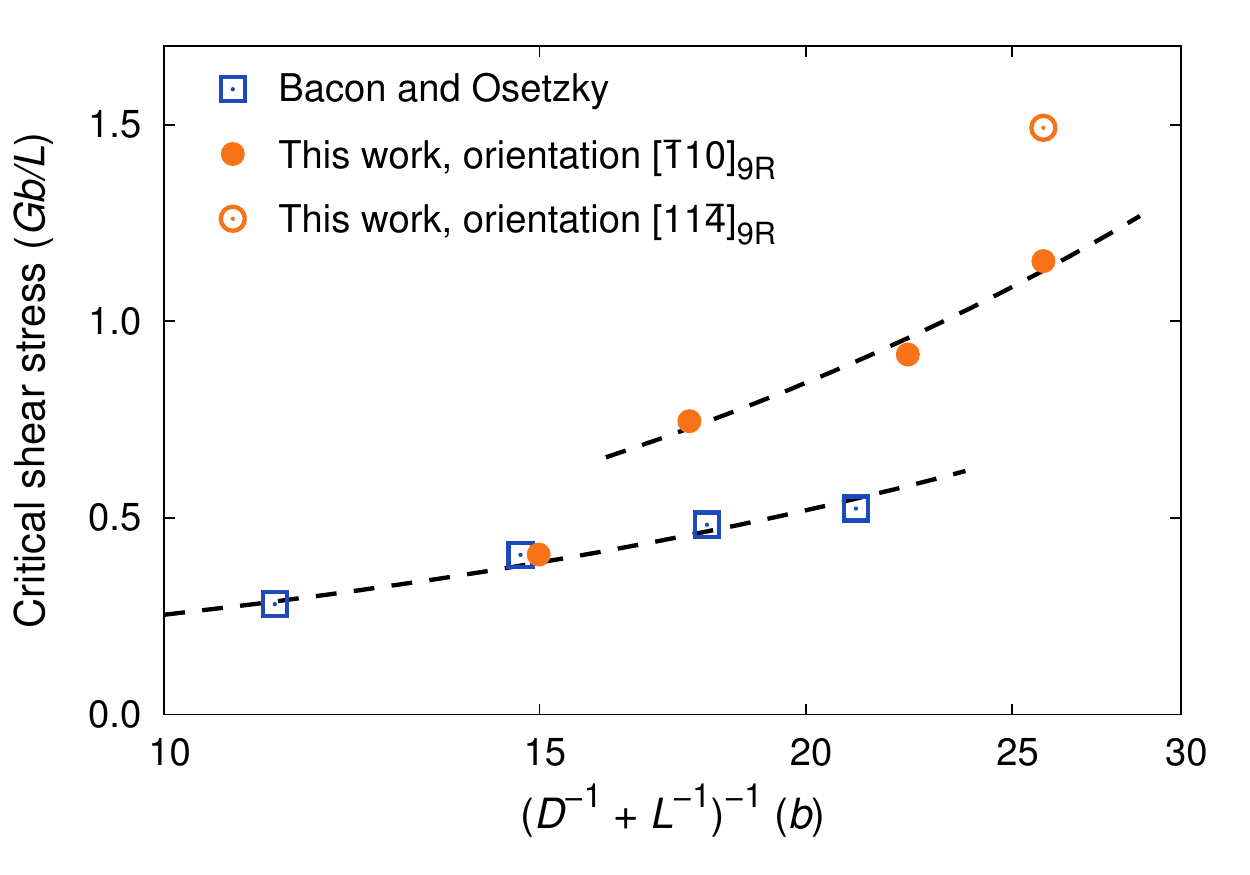}
  \caption{
    Critical shear stress as a function of harmonic average of $L$ and $D$ for MD simulations of screw dislocation-precipitate interactions performed here. Results are normalized to $\mu b/L$ to remove the dependence with $L$ and $\mu$ and facilitate comparison with earlier MD simulations. \cite{BacOse09}
  }
  \label{fig:stress}
\end{figure}

In \sect{sect:results_dislocations} it was shown that Cu precipitates can be a substantial source of resistance to screw dislocation motion in Fe--Cu alloys, and that for a given precipitate structure and orientation, hardening follows a DBH-type law. For that particular structure/orientation combination, the screw direction is a principal axis of the precipitate and there is no kinematic constraint to the dislocation traversing the obstacle; only the energetics of generating a dislocation segment in a different crystal structure are responsible for the observed hardening. Since we have obtained the hardening response from five precipitate sizes, including three different orientations, it is more informative to represent the values given in \tab{tab:data} as a function of the harmonic mean between $L$ and $D$. This is done in \fig{fig:stress}, which shows normalized critical stresses.

The figure reveals structure and orientation dependencies by placing different structures and/or orientations within distinct classes of behavior, as the dashed lines intended as guides to the eye reveal. Moreover, as the figure illustrates, our BCC Cu point is in excellent agreement with previous calculations by using a different interatomic potential Bacon and Osetsky, \cite{BacOse09} providing indirect verification for the dislocation-9R precipitate calculations.

As another useful verification exercise, we can confirm the interaction mechanism based on dislocation renucleation inside the precipitate (enabled by orientations alignment). To this end, more sophisticated models, such as the Russell-Brown modulus interaction model, \cite{RusBro72} that include the modulus mismatch between the matrix and the precipitate, must be invoked,
\begin{align}
  \tau_p = \frac{0.8\mu b}{L}\left(1-\frac{E_\Cu}{E}\right)^{\frac{3}{4}}
  \label{eq:rb}       
\end{align}
where $\mu$, $b$, and $L$ are defined as in \sect{sect:results_dislocations}, and $E_\Cu$ and $E$ are energies per unit length of the dislocation in precipitate and matrix, respectively. For screw dislocations, the ratio $E_\Cu/E$ is equivalent to \cite{Sto96}
\begin{align}
  \frac{E_\Cu}{E} = \frac{\mu_\Cu/\mu}{\log\left(r_{\infty}/r_0\right)}\left[\log{\left(\frac{r_\Cu}{r_0}\right)}
    +\log\left(\frac{r_{\infty}}{r_\Cu}\right)\right]
\end{align}
where $r_0$ and $r_{\infty}$ are the dislocation cutoff radius and outer cutoff radius used to compute the dislocation energy, $\mu_\Cu$ is the shear modulus of the precipitate (which we approximate as that of FCC-Cu), and $r_\Cu$ is the precipitate radius.

Equation~(\ref{eq:rb}) can be fitted to data from the interatomic potentials used in this work: $\mu=71.5\,\text{GPa}$, $\mu_\Cu=41.6\,\text{GPa}$ (again, rotated to the slip geometry used here from the nominal value of 76\,GPa), $r_0=2b$, $r_{\infty}=1000r_0$, and $r_\Cu=4b$. This yields a value of $\alpha=\left(1-E_\Cu/E\right)^{3/4}\approx\,0.5$, in good agreement with the MD analysis. This confirms that dislocation nucleation inside the precipitate is the dominant source of hardening, which is not surprising in light of the fact that we have chosen orientation relations that favor this mechanism over nucleation of interface partials. In any case, this observation runs counter to the conclusion extracted by Fine and Isheim \cite{FinIsh05} in their review of hardening models. Note that the values of $r_0$, $r_{\infty}$, and $r_\Cu$ are somewhat arbitrary, although due to their logarithmic dependence their effect is likely to be small. We note that our MD values are considerably smaller than those obtained using dislocation dynamics simulations, although these contain only a crude model of precipitate structure that may not yield the correct hardening. \cite{TakKur10}

In this context the observation of dislocation induced martensitic transformations in BCC Cu precipitates \cite{ShiChoKwo07, ShiKimJun08, BacOse09} that was alluded to above is also of interest.
\footnote{
  Note that the simulations in Refs.~\onlinecite{ShiChoKwo07, ShiKimJun08, BacOse09} employ other interatomic potentials than the present study.
}
In MD simulations carried out at 10\,K Shi and co-workers \cite{ShiChoKwo07, ShiKimJun08} observed the partial and largely reversible transformation of Cu atoms into a close-packed structure 9R-like structure in the presence of a screw dislocation for BCC precipitates with a diameter of at least 2.5\,nm (about 690 atoms). The fraction of transformed atoms was found to be strongly dependent on the distance between dislocation and precipitate surface. For distances above 1.5\,nm no transformation was observed up to the largest precipitate size considered (4\,nm equivalent to about 2,800 atoms), while for small distances more than 30\%\ of the Cu atoms transformed to a close-packed structure.

Bacon and Osetsky \cite{BacOse09} carried out a similar analysis as a function of temperature albeit for larger precipitates (6\,nm, about 9,700 atoms). In their simulations more than 50\%\ of the Cu atoms underwent a transformation at about 100\,K while this fraction dropped to about 10\%\ at 600\,K. These data are consistent with the results reported in the present paper, which provide additional insight into the origin of the observed behavior. The analysis of the temperature dependence of the BCC--9R transition (\sect{sect:results_bcc_9r_transition}) and its energetics (\sect{sect:results_thermodynamics}) provides a rationale for the strong temperature dependence observed by Bacon and Osetsky. It also resonates with their conclusion that the temperature dependence of the interaction of dislocations with Cu precipitates in $\alpha$-iron is more complex than for voids in the same material. \cite{BacOse09}

The correlation between the stress field of dislocations and the martensitic transformation in BCC Cu precipitates as well as the well-known shear instability of BCC Cu \cite{KraMarMet93, AckBacCal97, LiuWalGho05} suggest that the phase line between BCC and 9R Cu might not only be strongly temperature but also {\em shear} stress dependent.

\section{Conclusions}
\label{sect:conclusions}

We have studied Cu precipitates in Fe--Cu alloys using atomic scale simulations and empirical potentials. In terms of quantitative information, the main findings of this paper are ({\it i}) the polymorphic phase diagram of Cu precipitates in BCC Fe, and ({\it ii}) the strength (resistance to dislocation motion) of Cu precipitates with different structure. Finding ({\it i}) is validated by experimental data found in the literature (which provide shape, structure, and thermal information), whereas ({\it ii}) is verified against published MD simulations.

The BCC--9R transition is found to be strongly size and temperature dependent. This is explained in terms of the entropic stabilization of BCC versus close-packed Cu. While at high temperatures the transition is first-order it takes place without hysteresis at temperatures below approximately 300\,K. The latter behavior correlates with the maximum misorientation angle, which can be used to estimate the size of the nucleation barrier for the structural transformation. 

In addition we have extensively characterized the structure, morphology and strengthening effect of Cu precipitates. The present approach could conceivably be employed to determine the alloy composition and thermal treatments to tune the amount of precipitation hardening desired in an alloy.

\section*{Acknowledgments}

Part of this work was performed under the auspices of the U.S. Department of Energy by LLNL under Contract DE-AC52-07NA27344. P.E. acknowledges funding from the Swedish Research Council in the form of a Young Researcher grant, the European Research Council via a Marie Curie Career Integration Grant, and the {\em Area of Advance -- Materials Science} at Chalmers. B.S. acknowledges funding from the DOE-NE NEAMS program. Com\-puter time allocations by the National Energy Research Scientific Computing Center at Lawrence Berkeley National Laboratory and by the Swedish National Infrastructure for Computing are gratefully acknowledged.


\begin{thebibliography}{64}%
\makeatletter
\providecommand \@ifxundefined [1]{%
 \@ifx{#1\undefined}
}%
\providecommand \@ifnum [1]{%
 \ifnum #1\expandafter \@firstoftwo
 \else \expandafter \@secondoftwo
 \fi
}%
\providecommand \@ifx [1]{%
 \ifx #1\expandafter \@firstoftwo
 \else \expandafter \@secondoftwo
 \fi
}%
\providecommand \natexlab [1]{#1}%
\providecommand \enquote  [1]{``#1''}%
\providecommand \bibnamefont  [1]{#1}%
\providecommand \bibfnamefont [1]{#1}%
\providecommand \citenamefont [1]{#1}%
\providecommand \href@noop [0]{\@secondoftwo}%
\providecommand \href [0]{\begingroup \@sanitize@url \@href}%
\providecommand \@href[1]{\@@startlink{#1}\@@href}%
\providecommand \@@href[1]{\endgroup#1\@@endlink}%
\providecommand \@sanitize@url [0]{\catcode `\\12\catcode `\$12\catcode
  `\&12\catcode `\#12\catcode `\^12\catcode `\_12\catcode `\%12\relax}%
\providecommand \@@startlink[1]{}%
\providecommand \@@endlink[0]{}%
\providecommand \url  [0]{\begingroup\@sanitize@url \@url }%
\providecommand \@url [1]{\endgroup\@href {#1}{\urlprefix }}%
\providecommand \urlprefix  [0]{URL }%
\providecommand \Eprint [0]{\href }%
\providecommand \doibase [0]{http://dx.doi.org/}%
\providecommand \selectlanguage [0]{\@gobble}%
\providecommand \bibinfo  [0]{\@secondoftwo}%
\providecommand \bibfield  [0]{\@secondoftwo}%
\providecommand \translation [1]{[#1]}%
\providecommand \BibitemOpen [0]{}%
\providecommand \bibitemStop [0]{}%
\providecommand \bibitemNoStop [0]{.\EOS\space}%
\providecommand \EOS [0]{\spacefactor3000\relax}%
\providecommand \BibitemShut  [1]{\csname bibitem#1\endcsname}%
\let\auto@bib@innerbib\@empty
\bibitem [{\citenamefont {Odette}\ and\ \citenamefont
  {Lucas}(2001)}]{OdeLuc01}%
  \BibitemOpen
  \bibfield  {author} {\bibinfo {author} {\bibfnamefont {G.~R.}\ \bibnamefont
  {Odette}}\ and\ \bibinfo {author} {\bibfnamefont {G.~E.}\ \bibnamefont
  {Lucas}},\ }\href {\doibase 10.1007/s11837-001-0081-0} {\bibfield  {journal}
  {\bibinfo  {journal} {JOM}\ }\textbf {\bibinfo {volume} {53 (7)}},\ \bibinfo
  {pages} {18} (\bibinfo {year} {2001})}\BibitemShut {NoStop}%
\bibitem [{\citenamefont {Odette}(1983)}]{Ode83}%
  \BibitemOpen
  \bibfield  {author} {\bibinfo {author} {\bibfnamefont {G.~R.}\ \bibnamefont
  {Odette}},\ }\href@noop {} {\bibfield  {journal} {\bibinfo  {journal}
  {Scripta Metallurgica}\ }\textbf {\bibinfo {volume} {17}},\ \bibinfo {pages}
  {1183} (\bibinfo {year} {1983})}\BibitemShut {NoStop}%
\bibitem [{\citenamefont {Nakagawa}\ \emph {et~al.}(2007)\citenamefont
  {Nakagawa}, \citenamefont {Hori}, \citenamefont {Ohno}, \citenamefont
  {Ishikawa}, \citenamefont {Oshima}, \citenamefont {Kitagawa},\ and\
  \citenamefont {Iwase}}]{NakHorOhn07}%
  \BibitemOpen
  \bibfield  {author} {\bibinfo {author} {\bibfnamefont {S.}~\bibnamefont
  {Nakagawa}}, \bibinfo {author} {\bibfnamefont {F.}~\bibnamefont {Hori}},
  \bibinfo {author} {\bibnamefont {Ohno}}, \bibinfo {author} {\bibfnamefont
  {N.}~\bibnamefont {Ishikawa}}, \bibinfo {author} {\bibfnamefont
  {R.}~\bibnamefont {Oshima}}, \bibinfo {author} {\bibfnamefont
  {M.}~\bibnamefont {Kitagawa}}, \ and\ \bibinfo {author} {\bibfnamefont
  {A.}~\bibnamefont {Iwase}},\ }in\ \href@noop {} {\emph {\bibinfo {booktitle}
  {Materials Innovations for Next-Generation Nuclear Energy}}},\ Vol.\ \bibinfo
  {volume} {1043},\ \bibinfo {editor} {edited by\ \bibinfo {editor}
  {\bibfnamefont {R.}~\bibnamefont {Devanathan}}, \bibinfo {editor}
  {\bibfnamefont {R.}~\bibnamefont {Grimes}}, \bibinfo {editor} {\bibfnamefont
  {K.}~\bibnamefont {Yasuda}}, \bibinfo {editor} {\bibfnamefont
  {B.}~\bibnamefont {Uberuaga}}, \ and\ \bibinfo {editor} {\bibfnamefont
  {C.}~\bibnamefont {Meis}}}\ (\bibinfo {organization} {Materials Research
  Society},\ \bibinfo {year} {2007})\BibitemShut {NoStop}%
\bibitem [{\citenamefont {Othen}\ \emph {et~al.}(1991)\citenamefont {Othen},
  \citenamefont {Jenkins}, \citenamefont {Smith},\ and\ \citenamefont
  {Phythian}}]{OthJenSmi91}%
  \BibitemOpen
  \bibfield  {author} {\bibinfo {author} {\bibfnamefont {P.~J.}\ \bibnamefont
  {Othen}}, \bibinfo {author} {\bibfnamefont {M.~L.}\ \bibnamefont {Jenkins}},
  \bibinfo {author} {\bibfnamefont {G.~D.~W.}\ \bibnamefont {Smith}}, \ and\
  \bibinfo {author} {\bibfnamefont {W.~J.}\ \bibnamefont {Phythian}},\ }\href
  {\doibase 10.1080/09500839108215121} {\bibfield  {journal} {\bibinfo
  {journal} {Phil. Mag. Lett.}\ }\textbf {\bibinfo {volume} {64}},\ \bibinfo
  {pages} {383} (\bibinfo {year} {1991})}\BibitemShut {NoStop}%
\bibitem [{\citenamefont {Othen}\ \emph {et~al.}(1994)\citenamefont {Othen},
  \citenamefont {Jenkins},\ and\ \citenamefont {Smith}}]{OthJenSmi94}%
  \BibitemOpen
  \bibfield  {author} {\bibinfo {author} {\bibfnamefont {P.~J.}\ \bibnamefont
  {Othen}}, \bibinfo {author} {\bibfnamefont {M.~L.}\ \bibnamefont {Jenkins}},
  \ and\ \bibinfo {author} {\bibfnamefont {G.~D.~W.}\ \bibnamefont {Smith}},\
  }\href {\doibase 10.1080/01418619408242533} {\bibfield  {journal} {\bibinfo
  {journal} {Phil. Mag. A}\ }\textbf {\bibinfo {volume} {70}},\ \bibinfo
  {pages} {1} (\bibinfo {year} {1994})}\BibitemShut {NoStop}%
\bibitem [{Note1()}]{Note1}%
  \BibitemOpen
  \bibinfo {note} {The 3R structure is closely related to the face-centered
  cubic ground-state structure of Cu as it also exhibits a stacking sequence
  with threefold periodicity (see Sect.~\ref
  {sect:results_BCC_9R}).}\BibitemShut {Stop}%
\bibitem [{\citenamefont {Habibi-Bajguirani}\ and\ \citenamefont
  {Jenkins}(1996)}]{HabJen96}%
  \BibitemOpen
  \bibfield  {author} {\bibinfo {author} {\bibfnamefont {H.~R.}\ \bibnamefont
  {Habibi-Bajguirani}}\ and\ \bibinfo {author} {\bibfnamefont {M.~L.}\
  \bibnamefont {Jenkins}},\ }\href {\doibase 10.1080/095008396180786} {\
  \textbf {\bibinfo {volume} {73}},\ \bibinfo {pages} {155} (\bibinfo {year}
  {1996})}\BibitemShut {NoStop}%
\bibitem [{\citenamefont {Monzen}\ \emph {et~al.}(2000)\citenamefont {Monzen},
  \citenamefont {Jenkins},\ and\ \citenamefont {Sutton}}]{MonJenSut00}%
  \BibitemOpen
  \bibfield  {author} {\bibinfo {author} {\bibfnamefont {R.}~\bibnamefont
  {Monzen}}, \bibinfo {author} {\bibfnamefont {M.~L.}\ \bibnamefont {Jenkins}},
  \ and\ \bibinfo {author} {\bibfnamefont {A.~P.}\ \bibnamefont {Sutton}},\
  }\href@noop {} {\bibfield  {journal} {\bibinfo  {journal} {Phil. Mag. A}\
  }\textbf {\bibinfo {volume} {80}},\ \bibinfo {pages} {711} (\bibinfo {year}
  {2000})}\BibitemShut {NoStop}%
\bibitem [{\citenamefont {Lee}\ \emph {et~al.}(2007)\citenamefont {Lee},
  \citenamefont {Kim},\ and\ \citenamefont {Kim}}]{LeeKimKim07}%
  \BibitemOpen
  \bibfield  {author} {\bibinfo {author} {\bibfnamefont {T.-H.}\ \bibnamefont
  {Lee}}, \bibinfo {author} {\bibfnamefont {Y.-O.}\ \bibnamefont {Kim}}, \ and\
  \bibinfo {author} {\bibfnamefont {S.-J.}\ \bibnamefont {Kim}},\ }\href
  {\doibase 10.1080/14786430600909014} {\bibfield  {journal} {\bibinfo
  {journal} {Phil. Mag.}\ }\textbf {\bibinfo {volume} {87}},\ \bibinfo {pages}
  {209} (\bibinfo {year} {2007})}\BibitemShut {NoStop}%
\bibitem [{\citenamefont {Phythian}\ \emph {et~al.}(1992)\citenamefont
  {Phythian}, \citenamefont {Foreman}, \citenamefont {English}, \citenamefont
  {Buswell}, \citenamefont {Hetherington}, \citenamefont {Roberts},\ and\
  \citenamefont {Pizzini}}]{PhyForEng92}%
  \BibitemOpen
  \bibfield  {author} {\bibinfo {author} {\bibfnamefont {W.~J.}\ \bibnamefont
  {Phythian}}, \bibinfo {author} {\bibfnamefont {A.~J.~E.}\ \bibnamefont
  {Foreman}}, \bibinfo {author} {\bibfnamefont {C.~A.}\ \bibnamefont
  {English}}, \bibinfo {author} {\bibfnamefont {J.~T.}\ \bibnamefont
  {Buswell}}, \bibinfo {author} {\bibfnamefont {M.}~\bibnamefont
  {Hetherington}}, \bibinfo {author} {\bibfnamefont {K.}~\bibnamefont
  {Roberts}}, \ and\ \bibinfo {author} {\bibfnamefont {S.}~\bibnamefont
  {Pizzini}},\ }in\ \href {\doibase 10.1520/STP17866S} {\emph {\bibinfo
  {booktitle} {Effects of Radiation on Materials: 15th International
  Symposium}}},\ Vol.\ \bibinfo {volume} {STP1125},\ \bibinfo {editor} {edited
  by\ \bibinfo {editor} {\bibfnamefont {R.~E.}\ \bibnamefont {Stoller}},
  \bibinfo {editor} {\bibfnamefont {A.~S.}\ \bibnamefont {Kumar}}, \ and\
  \bibinfo {editor} {\bibfnamefont {D.~S.}\ \bibnamefont {Gelles}}}\ (\bibinfo
  {organization} {ASTM STP},\ \bibinfo {year} {1992})\ p.\ \bibinfo {pages}
  {1299}\BibitemShut {NoStop}%
\bibitem [{\citenamefont {Osetsky}\ \emph {et~al.}(1995)\citenamefont
  {Osetsky}, \citenamefont {Mikhin},\ and\ \citenamefont
  {Serra}}]{OseMikSer95}%
  \BibitemOpen
  \bibfield  {author} {\bibinfo {author} {\bibfnamefont {Y.~N.}\ \bibnamefont
  {Osetsky}}, \bibinfo {author} {\bibfnamefont {A.~G.}\ \bibnamefont {Mikhin}},
  \ and\ \bibinfo {author} {\bibfnamefont {A.}~\bibnamefont {Serra}},\ }\href
  {\doibase 10.1080/01418619508239930} {\bibfield  {journal} {\bibinfo
  {journal} {Phil. Mag. A}\ }\textbf {\bibinfo {volume} {72}},\ \bibinfo
  {pages} {361} (\bibinfo {year} {1995})}\BibitemShut {NoStop}%
\bibitem [{\citenamefont {Osetsky}\ and\ \citenamefont
  {Serra}(1996)}]{OseSer96}%
  \BibitemOpen
  \bibfield  {author} {\bibinfo {author} {\bibfnamefont {Y.~N.}\ \bibnamefont
  {Osetsky}}\ and\ \bibinfo {author} {\bibfnamefont {A.}~\bibnamefont
  {Serra}},\ }\href {\doibase 10.1080/01418619608242981} {\bibfield  {journal}
  {\bibinfo  {journal} {Phil. Mag. A}\ }\textbf {\bibinfo {volume} {73}},\
  \bibinfo {pages} {249} (\bibinfo {year} {1996})}\BibitemShut {NoStop}%
\bibitem [{\citenamefont {Osetsky}\ and\ \citenamefont
  {Serra}(1997)}]{OseSer97}%
  \BibitemOpen
  \bibfield  {author} {\bibinfo {author} {\bibfnamefont {Y.~N.}\ \bibnamefont
  {Osetsky}}\ and\ \bibinfo {author} {\bibfnamefont {A.}~\bibnamefont
  {Serra}},\ }\href {\doibase 10.1080/01418619708214013} {\bibfield  {journal}
  {\bibinfo  {journal} {Phil. Mag. A}\ }\textbf {\bibinfo {volume} {75}},\
  \bibinfo {pages} {1097} (\bibinfo {year} {1997})}\BibitemShut {NoStop}%
\bibitem [{\citenamefont {Ludwig}\ \emph {et~al.}(1998)\citenamefont {Ludwig},
  \citenamefont {Farkas}, \citenamefont {Pedraza},\ and\ \citenamefont
  {Schmauder}}]{LudFarPed98}%
  \BibitemOpen
  \bibfield  {author} {\bibinfo {author} {\bibfnamefont {M.}~\bibnamefont
  {Ludwig}}, \bibinfo {author} {\bibfnamefont {D.}~\bibnamefont {Farkas}},
  \bibinfo {author} {\bibfnamefont {D.}~\bibnamefont {Pedraza}}, \ and\
  \bibinfo {author} {\bibfnamefont {S.}~\bibnamefont {Schmauder}},\ }\href
  {\doibase 10.1088/0965-0393/6/1/003} {\bibfield  {journal} {\bibinfo
  {journal} {Model. Simul. Mater. Sci. Eng.}\ }\textbf {\bibinfo {volume}
  {6}},\ \bibinfo {pages} {19} (\bibinfo {year} {1998})}\BibitemShut {NoStop}%
\bibitem [{\citenamefont {Hu}\ \emph {et~al.}(1999)\citenamefont {Hu},
  \citenamefont {Li},\ and\ \citenamefont {Watanabe}}]{HuLiWat99}%
  \BibitemOpen
  \bibfield  {author} {\bibinfo {author} {\bibfnamefont {S.~Y.}\ \bibnamefont
  {Hu}}, \bibinfo {author} {\bibfnamefont {Y.~L.}\ \bibnamefont {Li}}, \ and\
  \bibinfo {author} {\bibfnamefont {K.}~\bibnamefont {Watanabe}},\ }\href
  {\doibase 10.1088/0965-0393/7/4/312} {\bibfield  {journal} {\bibinfo
  {journal} {Modelling Simul. Mater. Sci. Eng.}\ }\textbf {\bibinfo {volume}
  {7}},\ \bibinfo {pages} {641} (\bibinfo {year} {1999})}\BibitemShut {NoStop}%
\bibitem [{\citenamefont {Blackstock}\ and\ \citenamefont
  {Ackland}(2001)}]{BlaAck01}%
  \BibitemOpen
  \bibfield  {author} {\bibinfo {author} {\bibfnamefont {J.~J.}\ \bibnamefont
  {Blackstock}}\ and\ \bibinfo {author} {\bibfnamefont {G.~J.}\ \bibnamefont
  {Ackland}},\ }\href {\doibase 10.1080/01418610108217139} {\bibfield
  {journal} {\bibinfo  {journal} {Phil. Mag. A}\ }\textbf {\bibinfo {volume}
  {81}},\ \bibinfo {pages} {2127} (\bibinfo {year} {2001})}\BibitemShut
  {NoStop}%
\bibitem [{\citenamefont {Le~Bouar}(2001)}]{Bou01}%
  \BibitemOpen
  \bibfield  {author} {\bibinfo {author} {\bibfnamefont {Y.}~\bibnamefont
  {Le~Bouar}},\ }\href {\doibase 10.1016/S1359-6454(01)00178-1} {\bibfield
  {journal} {\bibinfo  {journal} {Acta Mater.}\ }\textbf {\bibinfo {volume}
  {49}},\ \bibinfo {pages} {2661} (\bibinfo {year} {2001})}\BibitemShut
  {NoStop}%
\bibitem [{\citenamefont {Shim}\ \emph {et~al.}(2007)\citenamefont {Shim},
  \citenamefont {Cho}, \citenamefont {Kwon}, \citenamefont {Kim},\ and\
  \citenamefont {Wirth}}]{ShiChoKwo07}%
  \BibitemOpen
  \bibfield  {author} {\bibinfo {author} {\bibfnamefont {J.-H.}\ \bibnamefont
  {Shim}}, \bibinfo {author} {\bibfnamefont {Y.~W.}\ \bibnamefont {Cho}},
  \bibinfo {author} {\bibfnamefont {S.~C.}\ \bibnamefont {Kwon}}, \bibinfo
  {author} {\bibfnamefont {W.~W.}\ \bibnamefont {Kim}}, \ and\ \bibinfo
  {author} {\bibfnamefont {B.~D.}\ \bibnamefont {Wirth}},\ }\href {\doibase
  10.1063/1.2429902} {\bibfield  {journal} {\bibinfo  {journal} {Appl. Phys.
  Lett.}\ }\textbf {\bibinfo {volume} {90}},\ \bibinfo {pages} {021906}
  (\bibinfo {year} {2007})}\BibitemShut {NoStop}%
\bibitem [{\citenamefont {Shim}\ \emph {et~al.}(2008)\citenamefont {Shim},
  \citenamefont {Kim}, \citenamefont {Jung}, \citenamefont {Cho}, \citenamefont
  {Hong},\ and\ \citenamefont {Wirth}}]{ShiKimJun08}%
  \BibitemOpen
  \bibfield  {author} {\bibinfo {author} {\bibfnamefont {J.-H.}\ \bibnamefont
  {Shim}}, \bibinfo {author} {\bibfnamefont {D.-I.}\ \bibnamefont {Kim}},
  \bibinfo {author} {\bibfnamefont {W.-S.}\ \bibnamefont {Jung}}, \bibinfo
  {author} {\bibfnamefont {Y.~W.}\ \bibnamefont {Cho}}, \bibinfo {author}
  {\bibfnamefont {K.~T.}\ \bibnamefont {Hong}}, \ and\ \bibinfo {author}
  {\bibfnamefont {B.~D.}\ \bibnamefont {Wirth}},\ }\href {\doibase
  doi:10.1063/1.3003083} {\bibfield  {journal} {\bibinfo  {journal} {J. Appl.
  Phys.}\ }\textbf {\bibinfo {volume} {104}},\ \bibinfo {pages} {083523}
  (\bibinfo {year} {2008})}\BibitemShut {NoStop}%
\bibitem [{\citenamefont {Bacon}\ and\ \citenamefont
  {Osetsky}(2009)}]{BacOse09}%
  \BibitemOpen
  \bibfield  {author} {\bibinfo {author} {\bibfnamefont {D.~J.}\ \bibnamefont
  {Bacon}}\ and\ \bibinfo {author} {\bibfnamefont {Y.~N.}\ \bibnamefont
  {Osetsky}},\ }\href {\doibase 10.1080/14786430903271377PII 917311877}
  {\bibfield  {journal} {\bibinfo  {journal} {Phil. Mag.}\ }\textbf {\bibinfo
  {volume} {89}},\ \bibinfo {pages} {3333} (\bibinfo {year}
  {2009})}\BibitemShut {NoStop}%
\bibitem [{\citenamefont {Sadigh}\ \emph {et~al.}(2012)\citenamefont {Sadigh},
  \citenamefont {Erhart}, \citenamefont {Stukowski}, \citenamefont {Caro},
  \citenamefont {Martinez},\ and\ \citenamefont {{Zepeda-Ruiz}}}]{SadErhStu12}%
  \BibitemOpen
  \bibfield  {author} {\bibinfo {author} {\bibfnamefont {B.}~\bibnamefont
  {Sadigh}}, \bibinfo {author} {\bibfnamefont {P.}~\bibnamefont {Erhart}},
  \bibinfo {author} {\bibfnamefont {A.}~\bibnamefont {Stukowski}}, \bibinfo
  {author} {\bibfnamefont {A.}~\bibnamefont {Caro}}, \bibinfo {author}
  {\bibfnamefont {E.}~\bibnamefont {Martinez}}, \ and\ \bibinfo {author}
  {\bibfnamefont {L.}~\bibnamefont {{Zepeda-Ruiz}}},\ }\href {\doibase
  10.1103/PhysRevB.85.184203} {\bibfield  {journal} {\bibinfo  {journal} {Phys.
  Rev. B}\ }\textbf {\bibinfo {volume} {85}},\ \bibinfo {pages} {184203}
  (\bibinfo {year} {2012})}\BibitemShut {NoStop}%
\bibitem [{\citenamefont {Mishin}\ \emph {et~al.}(2001)\citenamefont {Mishin},
  \citenamefont {Mehl}, \citenamefont {Papaconstantopoulos}, \citenamefont
  {Voter},\ and\ \citenamefont {Kress}}]{MisMehPap01}%
  \BibitemOpen
  \bibfield  {author} {\bibinfo {author} {\bibfnamefont {Y.}~\bibnamefont
  {Mishin}}, \bibinfo {author} {\bibfnamefont {M.~J.}\ \bibnamefont {Mehl}},
  \bibinfo {author} {\bibfnamefont {D.~A.}\ \bibnamefont
  {Papaconstantopoulos}}, \bibinfo {author} {\bibfnamefont {A.~F.}\
  \bibnamefont {Voter}}, \ and\ \bibinfo {author} {\bibfnamefont {J.~D.}\
  \bibnamefont {Kress}},\ }\href {\doibase 10.1103/PhysRevB.63.224106}
  {\bibfield  {journal} {\bibinfo  {journal} {Phys. Rev. B}\ }\textbf {\bibinfo
  {volume} {63}},\ \bibinfo {pages} {224106} (\bibinfo {year}
  {2001})}\BibitemShut {NoStop}%
\bibitem [{\citenamefont {Mendelev}\ \emph {et~al.}(2003)\citenamefont
  {Mendelev}, \citenamefont {Han}, \citenamefont {Srolovitz}, \citenamefont
  {Ackland}, \citenamefont {Sun},\ and\ \citenamefont {Asta}}]{MenHanSro03}%
  \BibitemOpen
  \bibfield  {author} {\bibinfo {author} {\bibfnamefont {M.~I.}\ \bibnamefont
  {Mendelev}}, \bibinfo {author} {\bibfnamefont {S.}~\bibnamefont {Han}},
  \bibinfo {author} {\bibfnamefont {D.~J.}\ \bibnamefont {Srolovitz}}, \bibinfo
  {author} {\bibfnamefont {G.~J.}\ \bibnamefont {Ackland}}, \bibinfo {author}
  {\bibfnamefont {D.~Y.}\ \bibnamefont {Sun}}, \ and\ \bibinfo {author}
  {\bibfnamefont {M.}~\bibnamefont {Asta}},\ }\href {\doibase
  10.1080/14786430310001613264} {\bibfield  {journal} {\bibinfo  {journal}
  {Phil. Mag.}\ }\textbf {\bibinfo {volume} {83}},\ \bibinfo {pages} {3977}
  (\bibinfo {year} {2003})}\BibitemShut {NoStop}%
\bibitem [{\citenamefont {Pasianot}\ and\ \citenamefont
  {Malerba}(2007)}]{PasMal07}%
  \BibitemOpen
  \bibfield  {author} {\bibinfo {author} {\bibfnamefont {R.~C.}\ \bibnamefont
  {Pasianot}}\ and\ \bibinfo {author} {\bibfnamefont {L.}~\bibnamefont
  {Malerba}},\ }\href {\doibase 10.1016/j.jnucmat.2006.09.008} {\bibfield
  {journal} {\bibinfo  {journal} {J. Nucl. Mater.}\ }\textbf {\bibinfo {volume}
  {360}},\ \bibinfo {pages} {118} (\bibinfo {year} {2007})}\BibitemShut
  {NoStop}%
\bibitem [{\citenamefont {Plimpton}(1995)}]{Pli95}%
  \BibitemOpen
  \bibfield  {author} {\bibinfo {author} {\bibfnamefont {S.}~\bibnamefont
  {Plimpton}},\ }\href {\doibase 10.1006/jcph.1995.1039} {\bibfield  {journal}
  {\bibinfo  {journal} {J. Comp. Phys.}\ }\textbf {\bibinfo {volume} {117}},\
  \bibinfo {pages} {1} (\bibinfo {year} {1995})}\BibitemShut {NoStop}%
\bibitem [{\citenamefont {Sadigh}\ and\ \citenamefont
  {Erhart}(2012)}]{SadErh12}%
  \BibitemOpen
  \bibfield  {author} {\bibinfo {author} {\bibfnamefont {B.}~\bibnamefont
  {Sadigh}}\ and\ \bibinfo {author} {\bibfnamefont {P.}~\bibnamefont
  {Erhart}},\ }\href {\doibase 10.1103/PhysRevB.86.134204} {\bibfield
  {journal} {\bibinfo  {journal} {Phys. Rev. B}\ }\textbf {\bibinfo {volume}
  {86}},\ \bibinfo {pages} {134204} (\bibinfo {year} {2012})}\BibitemShut
  {NoStop}%
\bibitem [{\citenamefont {Frenkel}\ and\ \citenamefont
  {Smit}(2001)}]{FreSmi01}%
  \BibitemOpen
  \bibfield  {author} {\bibinfo {author} {\bibfnamefont {D.}~\bibnamefont
  {Frenkel}}\ and\ \bibinfo {author} {\bibfnamefont {B.}~\bibnamefont {Smit}},\
  }\href@noop {} {\emph {\bibinfo {title} {Understanding Molecular
  Simulation}}}\ (\bibinfo  {publisher} {Academic Press},\ \bibinfo {address}
  {London},\ \bibinfo {year} {2001})\BibitemShut {NoStop}%
\bibitem [{Note2()}]{Note2}%
  \BibitemOpen
  \bibinfo {note} {It was shown in Refs.~\protect \rev@citealpnum {SadErhStu12,
  SadErh12} that the range of reasonable values for $\kappa $ extends over
  several orders of magnitude, rendering this choice uncritical.}\BibitemShut
  {Stop}%
\bibitem [{\citenamefont {Ackland}\ and\ \citenamefont
  {Jones}(2006)}]{AckJon06}%
  \BibitemOpen
  \bibfield  {author} {\bibinfo {author} {\bibfnamefont {G.~J.}\ \bibnamefont
  {Ackland}}\ and\ \bibinfo {author} {\bibfnamefont {A.~P.}\ \bibnamefont
  {Jones}},\ }\href {\doibase 10.1103/PhysRevB.73.054104} {\bibfield  {journal}
  {\bibinfo  {journal} {Phys. Rev. B}\ }\textbf {\bibinfo {volume} {73}},\
  \bibinfo {pages} {054104} (\bibinfo {year} {2006})}\BibitemShut {NoStop}%
\bibitem [{\citenamefont {Stukowski}(2010)}]{Stu10}%
  \BibitemOpen
  \bibfield  {author} {\bibinfo {author} {\bibfnamefont {A.}~\bibnamefont
  {Stukowski}},\ }\href {\doibase 10.1088/0965-0393/18/1/015012} {\bibfield
  {journal} {\bibinfo  {journal} {Model. Simul. Mater. Sci. Eng.}\ }\textbf
  {\bibinfo {volume} {18}},\ \bibinfo {pages} {015012} (\bibinfo {year}
  {2010})}\BibitemShut {NoStop}%
\bibitem [{Note3()}]{Note3}%
  \BibitemOpen
  \bibinfo {note} {The potentials employed here predict a solubility at
  700\protect \tmspace +\thinmuskip {.1667em}K of approximately 0.06\%. \cite
  {PasMal07} This value corresponds to the stability of the BCC Fe solid
  solution with respect to BCC Cu. Note that in order to obtain the solubility
  shown in standard phase diagrams one needs to consider the free energy
  balance between solid solutions of BCC Fe and FCC Cu. \cite {CarCarLop06a}
  The difference is, however, small and therefore neglected here.}\BibitemShut
  {Stop}%
\bibitem [{Note4()}]{Note4}%
  \BibitemOpen
  \bibinfo {note} {The 9R structure has also been observed e.g., at grain
  boundaries in silver \cite {ErnFinHof92} and gold \cite {MedFoiCoh01} as well
  as in single crystal copper wires. \cite {CheYanFan09}}\BibitemShut {NoStop}%
\bibitem [{\citenamefont {Kraft}\ \emph {et~al.}(1993)\citenamefont {Kraft},
  \citenamefont {Marcus}, \citenamefont {Methfessel},\ and\ \citenamefont
  {Scheffler}}]{KraMarMet93}%
  \BibitemOpen
  \bibfield  {author} {\bibinfo {author} {\bibfnamefont {T.}~\bibnamefont
  {Kraft}}, \bibinfo {author} {\bibfnamefont {P.~M.}\ \bibnamefont {Marcus}},
  \bibinfo {author} {\bibfnamefont {M.}~\bibnamefont {Methfessel}}, \ and\
  \bibinfo {author} {\bibfnamefont {M.}~\bibnamefont {Scheffler}},\ }\href
  {\doibase 10.1103/PhysRevB.48.5886} {\bibfield  {journal} {\bibinfo
  {journal} {Phys. Rev. B}\ }\textbf {\bibinfo {volume} {48}},\ \bibinfo
  {pages} {5886} (\bibinfo {year} {1993})}\BibitemShut {NoStop}%
\bibitem [{\citenamefont {Ackland}\ \emph {et~al.}(1997)\citenamefont
  {Ackland}, \citenamefont {Bacon}, \citenamefont {Calder},\ and\ \citenamefont
  {Harry}}]{AckBacCal97}%
  \BibitemOpen
  \bibfield  {author} {\bibinfo {author} {\bibfnamefont {G.~J.}\ \bibnamefont
  {Ackland}}, \bibinfo {author} {\bibfnamefont {D.~J.}\ \bibnamefont {Bacon}},
  \bibinfo {author} {\bibfnamefont {A.~F.}\ \bibnamefont {Calder}}, \ and\
  \bibinfo {author} {\bibfnamefont {T.}~\bibnamefont {Harry}},\ }\href
  {\doibase 10.1080/01418619708207198} {\bibfield  {journal} {\bibinfo
  {journal} {Phil. Mag. A}\ }\textbf {\bibinfo {volume} {75}},\ \bibinfo
  {pages} {713} (\bibinfo {year} {1997})}\BibitemShut {NoStop}%
\bibitem [{\citenamefont {Liu}\ \emph {et~al.}(2005)\citenamefont {Liu},
  \citenamefont {van~de Walle}, \citenamefont {Ghosh},\ and\ \citenamefont
  {Asta}}]{LiuWalGho05}%
  \BibitemOpen
  \bibfield  {author} {\bibinfo {author} {\bibfnamefont {J.~Z.}\ \bibnamefont
  {Liu}}, \bibinfo {author} {\bibfnamefont {A.}~\bibnamefont {van~de Walle}},
  \bibinfo {author} {\bibfnamefont {G.}~\bibnamefont {Ghosh}}, \ and\ \bibinfo
  {author} {\bibfnamefont {M.}~\bibnamefont {Asta}},\ }\href {\doibase
  10.1103/PhysRevB.72.144109} {\bibfield  {journal} {\bibinfo  {journal} {Phys.
  Rev. B}\ }\textbf {\bibinfo {volume} {72}},\ \bibinfo {pages} {144109}
  (\bibinfo {year} {2005})}\BibitemShut {NoStop}%
\bibitem [{Note5()}]{Note5}%
  \BibitemOpen
  \bibinfo {note} {For consistency with experiments, precipitate dimensions
  were measured parallel and perpendicular to the twin plane trace including
  only the 9R domains (that is excluding the Cu BCC shell, see Fig.~\ref
  {fig:viz_prec}). In this fashion, one obtains aspect ratios of 1.1, 1.3, and
  1.3 for the experimental precipitates in Fig.~2 of Ref.~\protect
  \rev@citealpnum {OthJenSmi91}, Fig.~4 of Ref.~\protect \rev@citealpnum
  {OthJenSmi94}, and Fig.~3 of Ref.~\protect \rev@citealpnum {LeeKimKim07},
  respectively.}\BibitemShut {Stop}%
\bibitem [{\citenamefont {Kajiwara}(1976)}]{Kaj76}%
  \BibitemOpen
  \bibfield  {author} {\bibinfo {author} {\bibfnamefont {S.}~\bibnamefont
  {Kajiwara}},\ }\href@noop {} {\bibfield  {journal} {\bibinfo  {journal}
  {Trans. Japan. Inst. Met.}\ }\textbf {\bibinfo {volume} {17}},\ \bibinfo
  {pages} {435} (\bibinfo {year} {1976})}\BibitemShut {NoStop}%
\bibitem [{Note6()}]{Note6}%
  \BibitemOpen
  \bibinfo {note} {The values for the fraction of atoms in BCC and FCC/HCP
  environments in Fig.~\ref {fig:md_quench} are lower than in Fig.~\ref
  {fig:trans_struct} because the Ackland-Jones analysis was carried out using
  instantaneous as opposed to time-averaged positions in order not to blur the
  temperature at which the structural transition occurs.}\BibitemShut {Stop}%
\bibitem [{Note7()}]{Note7}%
  \BibitemOpen
  \bibinfo {note} {The analysis of the potential energy has been carried out
  both by considering all atoms in the system (Fe matrix as well as Cu
  precipitate) and the Cu precipitate only with practically identical results.
  For the sake of clarity here we only show data for atoms in
  precipitates.}\BibitemShut {Stop}%
\bibitem [{Note8()}]{Note8}%
  \BibitemOpen
  \bibinfo {note} {This statement excludes very small precipitates ($N\protect
  \REV@lesssim \protect \tmspace +\thinmuskip {.1667em}500\protect \tmspace
  +\thinmuskip {.1667em}\protect \text {atoms}$), for which the wetting layer
  region comprises the entire precipitate. It is for the latter reason that in
  Fig.~\ref {fig:epot_trans}(a) the potential energy of small precipitates
  decreases rapidly with size.}\BibitemShut {Stop}%
\bibitem [{\citenamefont {Harry}\ and\ \citenamefont {Bacon}(2002)}]{HarBac02}%
  \BibitemOpen
  \bibfield  {author} {\bibinfo {author} {\bibfnamefont {T.}~\bibnamefont
  {Harry}}\ and\ \bibinfo {author} {\bibfnamefont {D.~J.}\ \bibnamefont
  {Bacon}},\ }\href@noop {} {\bibfield  {journal} {\bibinfo  {journal} {Acta
  Mater.}\ }\textbf {\bibinfo {volume} {50}},\ \bibinfo {pages} {195} (\bibinfo
  {year} {2002})}\BibitemShut {NoStop}%
\bibitem [{\citenamefont {Chen}\ \emph
  {et~al.}(2009{\natexlab{a}})\citenamefont {Chen}, \citenamefont {Kioussis},\
  and\ \citenamefont {Ghoniem}}]{CheKioGho09}%
  \BibitemOpen
  \bibfield  {author} {\bibinfo {author} {\bibfnamefont {Z.}~\bibnamefont
  {Chen}}, \bibinfo {author} {\bibfnamefont {N.}~\bibnamefont {Kioussis}}, \
  and\ \bibinfo {author} {\bibfnamefont {N.}~\bibnamefont {Ghoniem}},\
  }\href@noop {} {\bibfield  {journal} {\bibinfo  {journal} {Phys. Rev. B}\
  }\textbf {\bibinfo {volume} {80}},\ \bibinfo {pages} {184104} (\bibinfo
  {year} {2009}{\natexlab{a}})}\BibitemShut {NoStop}%
\bibitem [{\citenamefont {Lucas}(1993)}]{Luc93}%
  \BibitemOpen
  \bibfield  {author} {\bibinfo {author} {\bibfnamefont {G.~E.}\ \bibnamefont
  {Lucas}},\ }\href@noop {} {\bibfield  {journal} {\bibinfo  {journal} {J.
  Nucl. Mater.}\ }\textbf {\bibinfo {volume} {206}},\ \bibinfo {pages} {287}
  (\bibinfo {year} {1993})}\BibitemShut {NoStop}%
\bibitem [{Note9()}]{Note9}%
  \BibitemOpen
  \bibinfo {note} {It is worth noting that, although the DBH model is not
  successful in explaining many aspects of deformed irradiated materials (see,
  e.g., Ref.~\protect \rev@citealpnum {SinForTri97}), it provides a simple
  framework to show the connection between MD simulations of
  dislocation-precipitate interaction and continuum hardening
  laws.}\BibitemShut {Stop}%
\bibitem [{\citenamefont {Friedel}(1964)}]{Fri64}%
  \BibitemOpen
  \bibfield  {author} {\bibinfo {author} {\bibfnamefont {J.}~\bibnamefont
  {Friedel}},\ }\href@noop {} {\emph {\bibinfo {title} {Dislocations}}}\
  (\bibinfo  {publisher} {Pergamon},\ \bibinfo {address} {Oxford},\ \bibinfo
  {year} {1964})\ p.\ \bibinfo {pages} {379}\BibitemShut {NoStop}%
\bibitem [{\citenamefont {Robertson}\ \emph {et~al.}(2005)\citenamefont
  {Robertson}, \citenamefont {Beaudoin}, \citenamefont {Fadhalah},
  \citenamefont {Chun-Ming}, \citenamefont {Robach}, \citenamefont {Wirth},
  \citenamefont {Arsenlis}, \citenamefont {Ahn},\ and\ \citenamefont
  {Sofronis}}]{RobBeauFad05}%
  \BibitemOpen
  \bibfield  {author} {\bibinfo {author} {\bibfnamefont {I.~M.}\ \bibnamefont
  {Robertson}}, \bibinfo {author} {\bibfnamefont {A.}~\bibnamefont {Beaudoin}},
  \bibinfo {author} {\bibfnamefont {K.~A.}\ \bibnamefont {Fadhalah}}, \bibinfo
  {author} {\bibfnamefont {L.}~\bibnamefont {Chun-Ming}}, \bibinfo {author}
  {\bibfnamefont {J.}~\bibnamefont {Robach}}, \bibinfo {author} {\bibfnamefont
  {B.~D.}\ \bibnamefont {Wirth}}, \bibinfo {author} {\bibfnamefont
  {A.}~\bibnamefont {Arsenlis}}, \bibinfo {author} {\bibfnamefont
  {D.}~\bibnamefont {Ahn}}, \ and\ \bibinfo {author} {\bibfnamefont
  {P.}~\bibnamefont {Sofronis}},\ }\href@noop {} {\bibfield  {journal}
  {\bibinfo  {journal} {Mater. Sci. Eng. A}\ }\textbf {\bibinfo {volume}
  {400--401}},\ \bibinfo {pages} {245} (\bibinfo {year} {2005})}\BibitemShut
  {NoStop}%
\bibitem [{\citenamefont {Nedelcu}\ \emph {et~al.}(2000)\citenamefont
  {Nedelcu}, \citenamefont {Kizler}, \citenamefont {Schmauder},\ and\
  \citenamefont {Moldovan}}]{NedKizSch00}%
  \BibitemOpen
  \bibfield  {author} {\bibinfo {author} {\bibfnamefont {S.}~\bibnamefont
  {Nedelcu}}, \bibinfo {author} {\bibfnamefont {P.}~\bibnamefont {Kizler}},
  \bibinfo {author} {\bibfnamefont {S.}~\bibnamefont {Schmauder}}, \ and\
  \bibinfo {author} {\bibfnamefont {N.}~\bibnamefont {Moldovan}},\ }\href@noop
  {} {\bibfield  {journal} {\bibinfo  {journal} {Model. Simul. Mater. Sci.
  Eng.}\ }\textbf {\bibinfo {volume} {8}},\ \bibinfo {pages} {181} (\bibinfo
  {year} {2000})}\BibitemShut {NoStop}%
\bibitem [{\citenamefont {Marian}(2002)}]{Mar02}%
  \BibitemOpen
  \bibfield  {author} {\bibinfo {author} {\bibfnamefont {J.}~\bibnamefont
  {Marian}},\ }\emph {\bibinfo {title} {{Improved Understanding of Radiation
  Damage Modeling in Nuclear Materials}}},\ \href@noop {} {Ph.D. thesis},\
  \bibinfo  {school} {Universidad Politecnica de Madrid} (\bibinfo {year}
  {2002})\BibitemShut {NoStop}%
\bibitem [{\citenamefont {Bacon}\ and\ \citenamefont
  {Osetsky}(2004)}]{BacOse04}%
  \BibitemOpen
  \bibfield  {author} {\bibinfo {author} {\bibfnamefont {D.~J.}\ \bibnamefont
  {Bacon}}\ and\ \bibinfo {author} {\bibfnamefont {Y.~N.}\ \bibnamefont
  {Osetsky}},\ }\href@noop {} {\bibfield  {journal} {\bibinfo  {journal} {J.
  Nucl. Mater.}\ }\textbf {\bibinfo {volume} {{329-333}}},\ \bibinfo {pages}
  {1233} (\bibinfo {year} {2004})}\BibitemShut {NoStop}%
\bibitem [{\citenamefont {Kohler}\ \emph {et~al.}(2005)\citenamefont {Kohler},
  \citenamefont {Kizler},\ and\ \citenamefont {Schmauder}}]{KohKizSch05}%
  \BibitemOpen
  \bibfield  {author} {\bibinfo {author} {\bibfnamefont {C.}~\bibnamefont
  {Kohler}}, \bibinfo {author} {\bibfnamefont {P.}~\bibnamefont {Kizler}}, \
  and\ \bibinfo {author} {\bibfnamefont {S.}~\bibnamefont {Schmauder}},\
  }\href@noop {} {\bibfield  {journal} {\bibinfo  {journal} {Modelling Simul.
  Mater. Sci. Eng.}\ }\textbf {\bibinfo {volume} {13}},\ \bibinfo {pages} {35}
  (\bibinfo {year} {2005})}\BibitemShut {NoStop}%
\bibitem [{Note10()}]{Note10}%
  \BibitemOpen
  \bibinfo {note} {Values given correspond to those obtained after rotating the
  nominal shear modulus of 86\protect \tmspace +\thinmuskip {.1667em}GPa to the
  slip geometry used here.}\BibitemShut {Stop}%
\bibitem [{\citenamefont {Soisson}\ \emph {et~al.}(1996)\citenamefont
  {Soisson}, \citenamefont {Barbu},\ and\ \citenamefont
  {Martin}}]{SoiBarMar96}%
  \BibitemOpen
  \bibfield  {author} {\bibinfo {author} {\bibfnamefont {F.}~\bibnamefont
  {Soisson}}, \bibinfo {author} {\bibfnamefont {A.}~\bibnamefont {Barbu}}, \
  and\ \bibinfo {author} {\bibfnamefont {G.}~\bibnamefont {Martin}},\
  }\href@noop {} {\bibfield  {journal} {\bibinfo  {journal} {Acta Mater.}\
  }\textbf {\bibinfo {volume} {44}},\ \bibinfo {pages} {3789} (\bibinfo {year}
  {1996})}\BibitemShut {NoStop}%
\bibitem [{\citenamefont {Domain}\ \emph {et~al.}(1999)\citenamefont {Domain},
  \citenamefont {Becquart},\ and\ \citenamefont {Van~Duysen}}]{DomBecVan99}%
  \BibitemOpen
  \bibfield  {author} {\bibinfo {author} {\bibfnamefont {C.}~\bibnamefont
  {Domain}}, \bibinfo {author} {\bibfnamefont {C.~S.}\ \bibnamefont
  {Becquart}}, \ and\ \bibinfo {author} {\bibfnamefont {J.~C.}\ \bibnamefont
  {Van~Duysen}},\ }in\ \href@noop {} {\emph {\bibinfo {booktitle} {Multiscale
  Modelling of Materials}}},\ Vol.\ \bibinfo {volume} {538},\ \bibinfo {editor}
  {edited by\ \bibinfo {editor} {\bibfnamefont {V.~V.}\ \bibnamefont
  {Bulatov}}, \bibinfo {editor} {\bibfnamefont {T.}~\bibnamefont
  {{DiazdelaRubia}}}, \bibinfo {editor} {\bibfnamefont {R.}~\bibnamefont
  {Phillips}}, \bibinfo {editor} {\bibfnamefont {E.}~\bibnamefont {Kaxiras}}, \
  and\ \bibinfo {editor} {\bibfnamefont {N.}~\bibnamefont {Ghoniem}}}\
  (\bibinfo  {publisher} {Materials Research Society},\ \bibinfo {address}
  {Warrendale},\ \bibinfo {year} {1999})\ pp.\ \bibinfo {pages}
  {217--222}\BibitemShut {NoStop}%
\bibitem [{\citenamefont {Vincent}\ \emph {et~al.}(2008)\citenamefont
  {Vincent}, \citenamefont {Becquart}, \citenamefont {Pareige}, \citenamefont
  {Pareige},\ and\ \citenamefont {Domain}}]{VinBecPar08}%
  \BibitemOpen
  \bibfield  {author} {\bibinfo {author} {\bibfnamefont {E.}~\bibnamefont
  {Vincent}}, \bibinfo {author} {\bibfnamefont {C.~S.}\ \bibnamefont
  {Becquart}}, \bibinfo {author} {\bibfnamefont {C.}~\bibnamefont {Pareige}},
  \bibinfo {author} {\bibfnamefont {P.}~\bibnamefont {Pareige}}, \ and\
  \bibinfo {author} {\bibfnamefont {C.}~\bibnamefont {Domain}},\ }\href
  {\doibase 10.1016/j.jnucmat.2007.06.016} {\bibfield  {journal} {\bibinfo
  {journal} {J. Nucl. Mater.}\ }\textbf {\bibinfo {volume} {373}},\ \bibinfo
  {pages} {387} (\bibinfo {year} {2008})}\BibitemShut {NoStop}%
\bibitem [{\citenamefont {Russell}\ and\ \citenamefont
  {Brown}(1972)}]{RusBro72}%
  \BibitemOpen
  \bibfield  {author} {\bibinfo {author} {\bibfnamefont {K.~C.}\ \bibnamefont
  {Russell}}\ and\ \bibinfo {author} {\bibfnamefont {L.~M.}\ \bibnamefont
  {Brown}},\ }\href@noop {} {\bibfield  {journal} {\bibinfo  {journal} {Acta
  Metall.}\ }\textbf {\bibinfo {volume} {20}},\ \bibinfo {pages} {969}
  (\bibinfo {year} {1972})}\BibitemShut {NoStop}%
\bibitem [{\citenamefont {Stoller}(1996)}]{Sto96}%
  \BibitemOpen
  \bibfield  {author} {\bibinfo {author} {\bibfnamefont {R.~E.}\ \bibnamefont
  {Stoller}},\ }in\ \href@noop {} {\emph {\bibinfo {booktitle} {Effects of
  Radiation on Materials: 17$^{th}$ International Symposium}}},\ Vol.\ \bibinfo
  {volume} {ASTM STP 1270},\ \bibinfo {editor} {edited by\ \bibinfo {editor}
  {\bibfnamefont {D.~S.}\ \bibnamefont {Gelles}}, \bibinfo {editor}
  {\bibfnamefont {R.~K.}\ \bibnamefont {Nanstad}}, \bibinfo {editor}
  {\bibfnamefont {A.~S.}\ \bibnamefont {Kumar}}, \ and\ \bibinfo {editor}
  {\bibfnamefont {E.~A.}\ \bibnamefont {Little}}}\ (\bibinfo {organization}
  {American Society for Testing and Materials},\ \bibinfo {year}
  {1996})\BibitemShut {NoStop}%
\bibitem [{\citenamefont {Fine}\ and\ \citenamefont {Isheim}(2005)}]{FinIsh05}%
  \BibitemOpen
  \bibfield  {author} {\bibinfo {author} {\bibfnamefont {M.~E.}\ \bibnamefont
  {Fine}}\ and\ \bibinfo {author} {\bibfnamefont {D.}~\bibnamefont {Isheim}},\
  }\href {\doibase 10.1016/j.scriptamat.2005.02.034} {\bibfield  {journal}
  {\bibinfo  {journal} {Scripta Mater.}\ }\textbf {\bibinfo {volume} {53}},\
  \bibinfo {pages} {115} (\bibinfo {year} {2005})}\BibitemShut {NoStop}%
\bibitem [{\citenamefont {Takahashi}\ and\ \citenamefont
  {Kurata}(2010)}]{TakKur10}%
  \BibitemOpen
  \bibfield  {author} {\bibinfo {author} {\bibfnamefont {A.}~\bibnamefont
  {Takahashi}}\ and\ \bibinfo {author} {\bibfnamefont {K.}~\bibnamefont
  {Kurata}},\ }\href {\doibase 10.1088/1757-899X/10/1/012081} {\bibfield
  {journal} {\bibinfo  {journal} {IOP Conf. Ser.: Mater. Sci. Eng.}\ }\textbf
  {\bibinfo {volume} {10}},\ \bibinfo {pages} {012081} (\bibinfo {year}
  {2010})}\BibitemShut {NoStop}%
\bibitem [{Note11()}]{Note11}%
  \BibitemOpen
  \bibinfo {note} {Note that the simulations in Refs.~\protect \rev@citealpnum
  {ShiChoKwo07, ShiKimJun08, BacOse09} employ other interatomic potentials than
  the present study.}\BibitemShut {Stop}%
\bibitem [{\citenamefont {Caro}\ \emph {et~al.}(2006)\citenamefont {Caro},
  \citenamefont {Caro}, \citenamefont {Lopasso}, \citenamefont {Turchi},\ and\
  \citenamefont {Farkas}}]{CarCarLop06a}%
  \BibitemOpen
  \bibfield  {author} {\bibinfo {author} {\bibfnamefont {A.}~\bibnamefont
  {Caro}}, \bibinfo {author} {\bibfnamefont {M.}~\bibnamefont {Caro}}, \bibinfo
  {author} {\bibfnamefont {E.~M.}\ \bibnamefont {Lopasso}}, \bibinfo {author}
  {\bibfnamefont {P.~E.~A.}\ \bibnamefont {Turchi}}, \ and\ \bibinfo {author}
  {\bibfnamefont {D.}~\bibnamefont {Farkas}},\ }\href@noop {} {\bibfield
  {journal} {\bibinfo  {journal} {J. Nucl. Mater.}\ }\textbf {\bibinfo {volume}
  {349}},\ \bibinfo {pages} {317} (\bibinfo {year} {2006})}\BibitemShut
  {NoStop}%
\bibitem [{\citenamefont {Ernst}\ \emph {et~al.}(1992)\citenamefont {Ernst},
  \citenamefont {Finnis}, \citenamefont {Hofmann}, \citenamefont {Muschik},
  \citenamefont {Sch\"onberger}, \citenamefont {Wolf},\ and\ \citenamefont
  {Methfessel}}]{ErnFinHof92}%
  \BibitemOpen
  \bibfield  {author} {\bibinfo {author} {\bibfnamefont {F.}~\bibnamefont
  {Ernst}}, \bibinfo {author} {\bibfnamefont {M.~W.}\ \bibnamefont {Finnis}},
  \bibinfo {author} {\bibfnamefont {D.}~\bibnamefont {Hofmann}}, \bibinfo
  {author} {\bibfnamefont {T.}~\bibnamefont {Muschik}}, \bibinfo {author}
  {\bibfnamefont {U.}~\bibnamefont {Sch\"onberger}}, \bibinfo {author}
  {\bibfnamefont {U.}~\bibnamefont {Wolf}}, \ and\ \bibinfo {author}
  {\bibfnamefont {M.}~\bibnamefont {Methfessel}},\ }\href {\doibase
  10.1103/PhysRevLett.69.620} {\bibfield  {journal} {\bibinfo  {journal} {Phys.
  Rev. Lett.}\ }\textbf {\bibinfo {volume} {69}},\ \bibinfo {pages} {620}
  (\bibinfo {year} {1992})}\BibitemShut {NoStop}%
\bibitem [{\citenamefont {Medlin}\ \emph {et~al.}(2001)\citenamefont {Medlin},
  \citenamefont {Foiles},\ and\ \citenamefont {Cohen}}]{MedFoiCoh01}%
  \BibitemOpen
  \bibfield  {author} {\bibinfo {author} {\bibfnamefont {D.}~\bibnamefont
  {Medlin}}, \bibinfo {author} {\bibfnamefont {S.}~\bibnamefont {Foiles}}, \
  and\ \bibinfo {author} {\bibfnamefont {D.}~\bibnamefont {Cohen}},\ }\href
  {\doibase 10.1016/S1359-6454(01)00284-1} {\bibfield  {journal} {\bibinfo
  {journal} {Acta Mater.}\ }\textbf {\bibinfo {volume} {49}},\ \bibinfo {pages}
  {3689} (\bibinfo {year} {2001})}\BibitemShut {NoStop}%
\bibitem [{\citenamefont {Chen}\ \emph
  {et~al.}(2009{\natexlab{b}})\citenamefont {Chen}, \citenamefont {Yan},\ and\
  \citenamefont {Fan}}]{CheYanFan09}%
  \BibitemOpen
  \bibfield  {author} {\bibinfo {author} {\bibfnamefont {J.}~\bibnamefont
  {Chen}}, \bibinfo {author} {\bibfnamefont {W.}~\bibnamefont {Yan}}, \ and\
  \bibinfo {author} {\bibfnamefont {X.-H.}\ \bibnamefont {Fan}},\ }\href
  {\doibase 10.1016/S1003-6326(08)60236-8} {\bibfield  {journal} {\bibinfo
  {journal} {Transactions of Nonferrous Metals Society of China}\ }\textbf
  {\bibinfo {volume} {19}},\ \bibinfo {pages} {108} (\bibinfo {year}
  {2009}{\natexlab{b}})}\BibitemShut {NoStop}%
\bibitem [{\citenamefont {Singh}\ \emph {et~al.}(1997)\citenamefont {Singh},
  \citenamefont {Foreman},\ and\ \citenamefont {Trinkaus}}]{SinForTri97}%
  \BibitemOpen
  \bibfield  {author} {\bibinfo {author} {\bibfnamefont {B.~N.}\ \bibnamefont
  {Singh}}, \bibinfo {author} {\bibfnamefont {A.~J.~E.}\ \bibnamefont
  {Foreman}}, \ and\ \bibinfo {author} {\bibfnamefont {H.}~\bibnamefont
  {Trinkaus}},\ }\href@noop {} {\bibfield  {journal} {\bibinfo  {journal} {J.
  Nucl. Mater.}\ }\textbf {\bibinfo {volume} {249}},\ \bibinfo {pages} {103}
  (\bibinfo {year} {1997})}\BibitemShut {NoStop}%
\end{thebibliography}

%

\end{document}